\documentclass{JHEP3}
\usepackage{graphicx}
\usepackage{latexsym}
\usepackage{amsmath}
\usepackage{amsfonts}
\usepackage{amssymb}
\newcommand{\be}{\begin{equation}}
\newcommand{\ee}{\end{equation}}
\newcommand{\ben}{\begin{eqnarray}}
\newcommand{\een}{\end{eqnarray}}
\newcommand{\bes}{\begin{subequations}}
\newcommand{\ees}{\end{subequations}}
\newcommand{\bb}{\bibitem}

\newcommand{\nn}{\nonumber\\}

\title{Scalar fields, bent branes, and RG flow}

\author{Dionisio Bazeia,$^a$ Francisco A. Brito,$^b$ and Laercio Losano\,$^a$\\
$^a$Departamento de F\'\i sica, Universidade Federal da
Para\'\i ba\\
Caixa Postal 5008, 58051-970 Jo\~ao Pessoa, Para\'\i ba, Brazil\\
$^b$Departamento de F\'\i sica, Universidade Federal de Campina
Grande\\ Caixa Postal 10071, 58109-970  Campina Grande, Para\'\i ba,
Brazil\\
\\
{\rm E-mails:\,}{bazeia@fisica.ufpb.br, fabrito@df.ufcg.edu.br,
losano@fisica.ufpb.br}}

\abstract{This work deals with braneworld scenarios driven by 
real scalar fields with standard dynamics. We show how the
first-order formalism which exists in the case of four dimensional
Minkowski space-time can be extended to de Sitter or anti-de Sitter
geometry in the presence of several real scalar fields.
We illustrate the results with some examples, and we
take advantage of our findings to investigate renormalization group
flow. We have found symmetric brane solutions with four-dimensional
anti-de Sitter geometry whose holographically dual field theory
exhibits a weakly coupled regime at high energy.
 \\

\vspace{.7cm}

}

\maketitle

\begin{document}

\section{Introduction}

In the present work, we focus attention on braneworld models
described in five-dimensional space-time with warped geometry
involving a single infinitely large extra dimension, as firstly
introduced in Ref.~{\cite{rs2}}. Following the original work of
Randall and Sundrum, the braneworld scenario that we consider
appears with five-dimensional AdS (or even Minkowski) geometry, and
the 3-brane is now allowed to have four-dimensional space-time with AdS,
Minkowski, or dS geometry \cite{gw,s,f,g1,g2,c1,c2,klp,kr,bfg,fn,celi,bcy,cv_lamb,bbn}.

The model describes five-dimensional gravity in the presence of
dynamical bulk scalar fields. Our motivation arises from the
investigations \cite{s,f,bfg,fn,celi,bglm,abl2006,st}, which consider
the possibility of finding first-order differential equations which
solve the corresponding equations of motion.  This is technically
important, since it directly contributes to simplify investigations,
and to open new scenarios. The present study is connected with the
former work \cite{abl2006} and inspired in similar calculations done
by some of us in cosmology \cite{bglm}. It is interesting to see
that, although supersymmetry seem to be incompatible with dS
geometry, even in this case we have found a way to write first-order
equations which solve the equations of motion of the original system
\cite{abl2006}. See also Ref.~{\cite{st2}} for another procedure,
based on the Hamilton-Jacobi formalism.

The investigations focus on Einstein's equation and the equations of
motion for the scalar fields in a very direct way. We consider
models described by real scalar fields in five-dimensional
space-time with anti-de Sitter (AdS), or Minkowski ($\mathbb{M}$)
geometry, which engenders a single extra dimension and generic four
dimensional space-time with AdS, Minkowski, or dS geometry. The
scalar fields are described with standard dynamics, and we follow a
very specific route, set forward in \cite{bglm}, in which we use the
potential of the scalar field to infer how the warp factor depends
on the extra dimension.

The power of the method that we develop is related to an important
simplification, which leads to models governed by scalar field
potential of very specific form, depending on two new functions,
$W=W(\phi)$ and $Z=Z(\phi).$ As we show below, we relate the
function $W(\phi)$ to the warp factor, and this leads to scenarios
where the scalar field may be connected with $W$ and $Z,$ unveiling
a new route to investigate the subject.

We illustrate our findings with several examples of current
interest, described by two real scalar fields, and we take advantage
of our findings to investigate renormalization group flow for flat
branes \cite{skenderis,freedman_pilch,s} and ``bent'' (or
``curved'') branes \cite{dallagata}. We have found symmetric bent
brane solutions with four-dimensional AdS geometry whose
holographically dual field theory exhibits a weakly coupled regime
at high energy. Such bent  branes may give a dual gravitational
description of RG flows in supersymmetric field theories living in
the curved spacetime of the brane world-volume \cite{dallagata}.
Furthermore, such $AdS_4$ brane theories exhibit improved infrared
behavior. Other examples include {\it asymmetric branes} that are
asymptotically $\mathbb{M}_5-AdS_5$ spaces. We find that in these
theories there exists a `natural' UV cut-off on the running coupling
in the Minkowski ($\mathbb{M}_5$) side.

The paper is organized as follows. In Sec.~\ref{one} we look for
flat and curved thick 3-brane solutions in a five-dimensional theory
of gravity coupled with one scalar field. In Sec.~\ref{two} we
extend the earlier analysis to many scalar fields. In
Sec.~\ref{gravity} we investigate the localization of gravity for
the solutions we find, and point out some issues in obtaining the
calculations analytically. In Sec.~\ref{RGflow} we study the
implications of the renormalization group flow for the dual field
theory on the boundary of the five-dimensional spacetime. We make
our final considerations in Sec.~\ref{conclu}.

\section{One scalar field }
\label{one}
The models that we investigate is described by a theory of
five-dimensional gravity coupled to scalar fields governed by the
following action \be\label{model}
S=\int\,d^4xdy\;{\sqrt{|g|}\;\left(-\frac14\,R+{\cal
L}(\phi,\partial_i\phi)\right)}, \ee where $\phi$ stands for a real
scalar field and we are using ${4\pi G}=1.$ These theories with one
or many scalar fields can simulate true five-dimensional
supergravity theories under certain consistent truncations --- see
Refs.~\cite{celi,st} for further discussions. The line element
$ds^2_5$ of the five-dimensional space-time can be written as
\be\label{metric} ds^2_5=g_{ij}dx^idx^j=e^{2A}ds^2_4-dy^2, \ee for
$i,j=0,1,...,4.$ Also, $ds^2_4$ represents the line element of the
four-dimensional space-time, which can have the form \bes \ben
ds^2_4=dt^2-e^{2\sqrt{\Lambda}t}(dx_1^2+dx_2^2+dx_3^2),
\\
ds^2_4=e^{-2\sqrt{\Lambda}x_3}(dt^2-dx_1^2-dx_2^2)-dx_3^2, \een\ees
for dS or AdS geometry, respectively. Here $e^{2A}$ is the warp
factor and $\Lambda$ represents the cosmological constant of the
four-dimensional space-time; the limit $\Lambda\to0$ leads to the
line element \be ds^2_5=e^{2A}\eta_{\mu\nu}dx^\mu dx^\nu-dy^2, \ee
where $\eta_{\mu\nu},\;\mu,\nu=0,1,2,3,$ describes Minkowski
geometry. The scalar field dynamics is governed by the Lagrangian
density \be {\cal
L}=\frac12\,g_{ij}\,\partial^i\phi\,\partial^j\phi-V,\ee where
$V=V(\phi)$ represents the potential, which specifies the model to
be considered.

\subsection{Flat branes}

Let us first focus on flat (or Minkowski) branes, i.e., the cases
with $\Lambda=0$. As usual, we suppose that both $A$ and $\phi$ are
static, and depend only on the extra dimension, that is, we set
$A=A(y)$ and $\phi=\phi(y).$ In this case the equation of motion for
the scalar field has the form \be\label{ephi}
\phi^{\prime\prime}+4A^\prime\phi^\prime=V_\phi, \ee where prime
denotes derivative with respect to $y,$ and $V_\phi=dV/d\phi.$ The
Einstein's equation with Minkowski four-dimensional geometry gives
\bes\label{ee0} \ben
A^{\prime\prime}=-\frac23\phi^{\prime2}\label{ee02},
\\
A^{\prime2}=\frac16\phi^{\prime2}-\frac13V(\phi)\label{ee01}. \een
\ees To get to the first-order formalism \cite{s,f,g1,g2}, we
introduce another function, $W=W(\phi),$ which can be viewed as a
superpotential in supergravity extensions. By writing the
first-order equation \be\label{A1} A^\prime=-\frac13 W, \ee and
using the equation (\ref{ee02}) we get to \be\label{phi1}
\phi^\prime=\frac12W_\phi, \ee and now the potential in
Eq.~(\ref{ee01}) has the form \be\label{p1}
V=\frac18W^2_\phi-\frac13 W^2. \ee It is not difficult to show that
Eqs.~(\ref{A1}) and (\ref{phi1}) solve Eqs.~(\ref{ephi}) and
(\ref{ee0}) for the above potential (\ref{p1}). We can also change
$W\to-W$ to get another possibility without changing the potential.
This result is very interesting, since it simplifies the calculation
significantly. As one knows, it was already obtained in former works
\cite{s,f}.

In the following we illustrate the procedure with several flat brane
examples. The first example is given by the superpotential \cite{V,p}
\be\label{wsech}
W=2a\arctan(\sinh(b\phi),
\ee
which gives the following scalar potential
\be V(\phi)=\frac12 a^2b^2{\rm
sech}^2(b\phi)-\frac43 a^2\arctan^2\left(\sinh^2(b\phi)\right).
\ee
The solution of the first-order equations is
\be
\phi(y)=\pm\frac{1}{b}{\rm arcsinh}(ab^2y),
\ee
and
\be\label{ay1}
A(y)=\frac{1}{3b^2}\ln(1+a^2b^4y^2)-\frac23 a\;y\;\arctan(ab^2y).
\ee

\begin{figure}[!ht]
\vspace{.3cm}
\includegraphics[{height=7.5cm,width=7cm,angle=-90}]{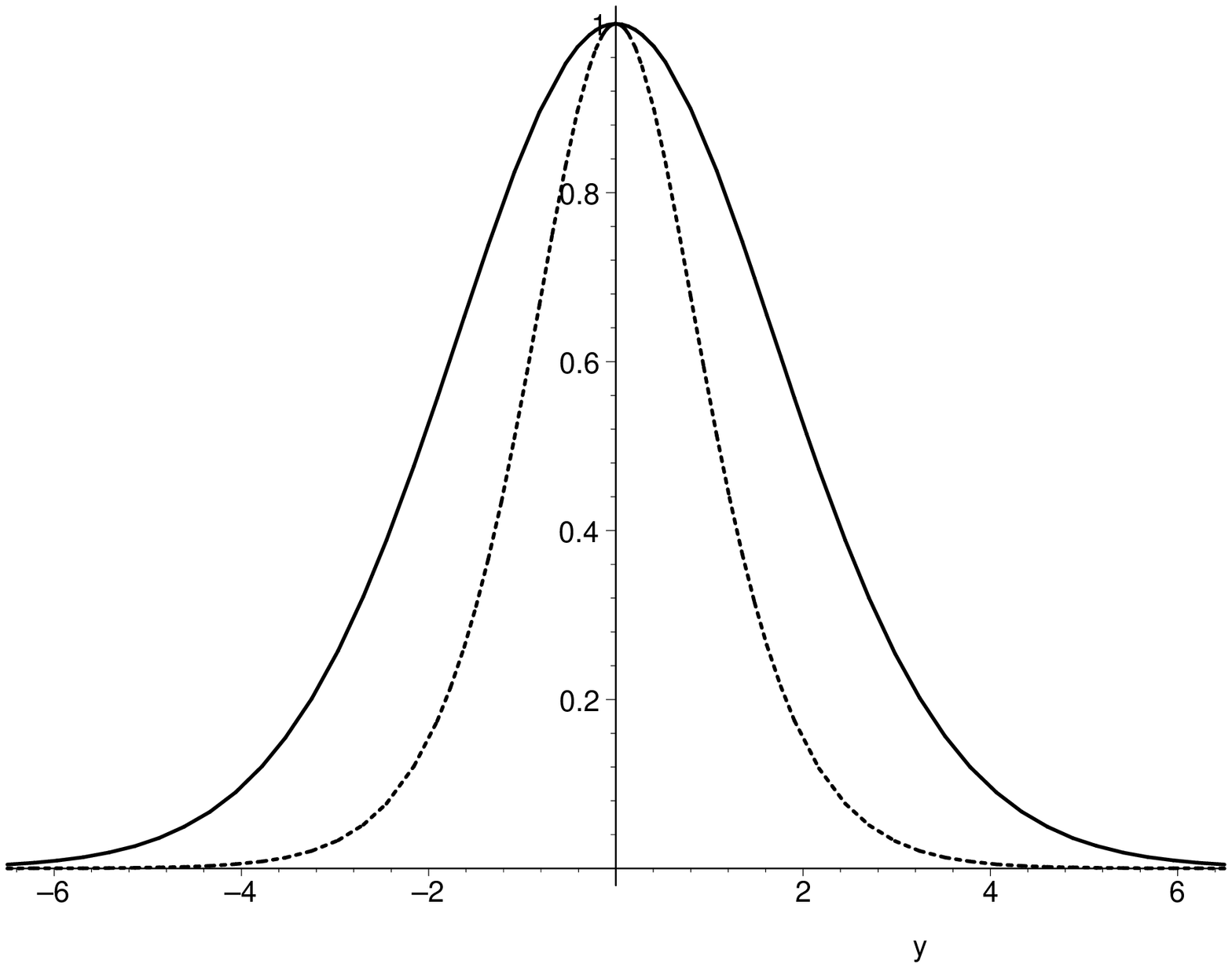}
\includegraphics[{height=7.5cm,width=7cm,angle=-90}]{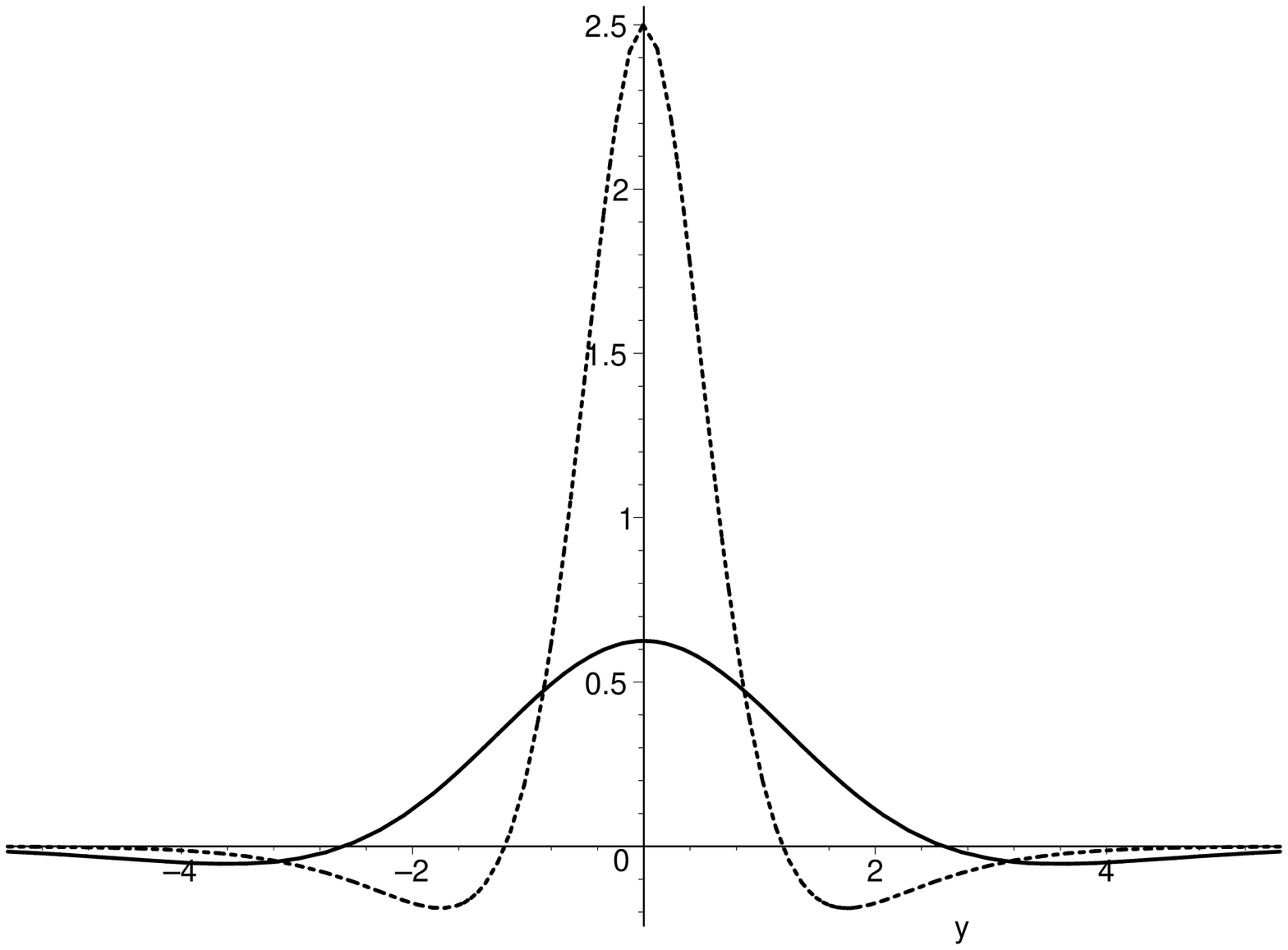}
\vspace{0.3cm}
\caption{Warp factor (left panel) for $a=1$, and
$b=1/2$ (solid line), $b=1$ (dashed line), and the corresponding
energy densities (right panel) for the scalar fields in curved
space-time for the model described by Eq.~(2.11).}\label{fig_wsech}
\end{figure}
\begin{figure}[!ht]
\vspace{.3cm}
\includegraphics[{height=7.5cm,width=7cm,angle=-90}]{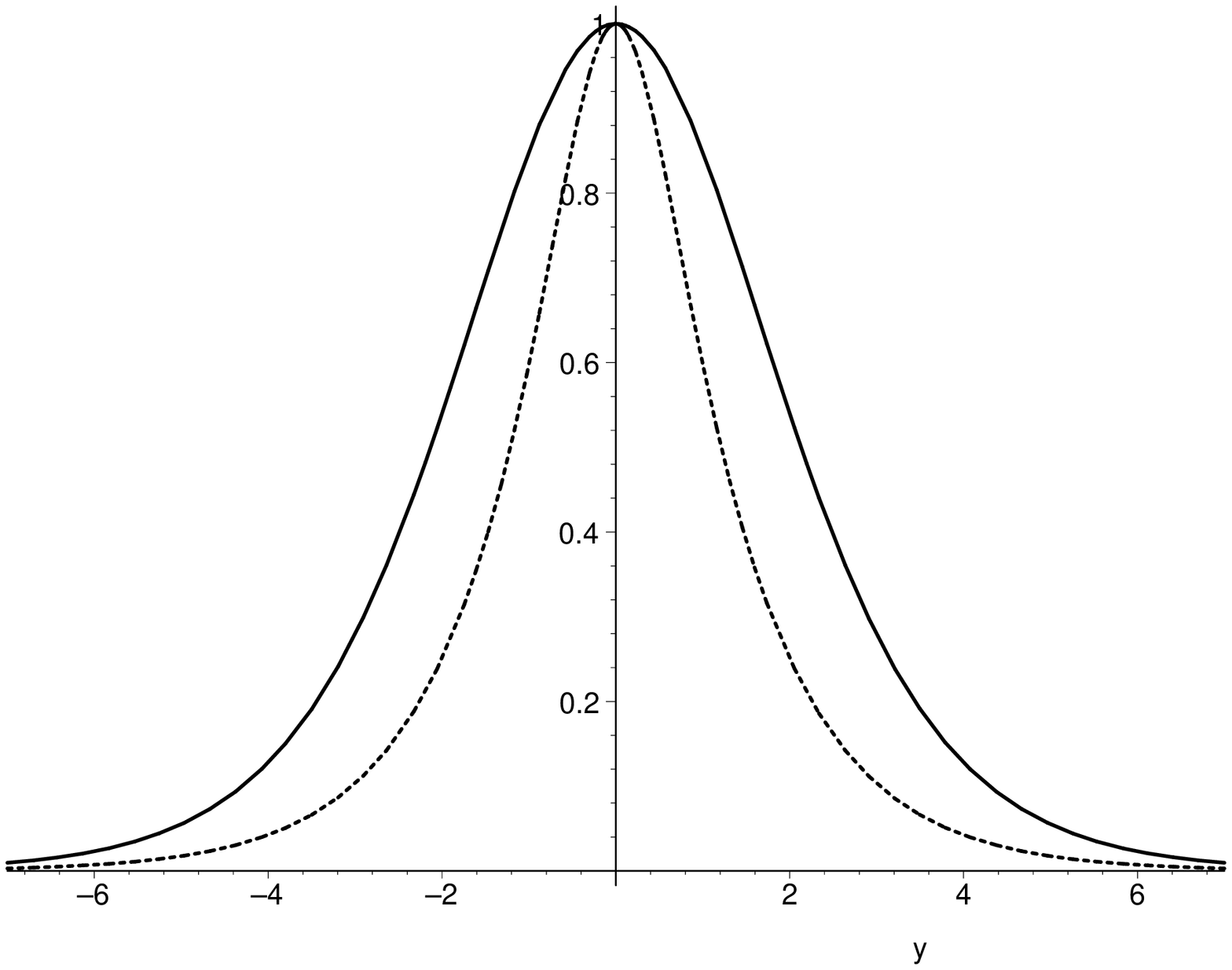}
\includegraphics[{height=7.5cm,width=7cm,angle=-90}]{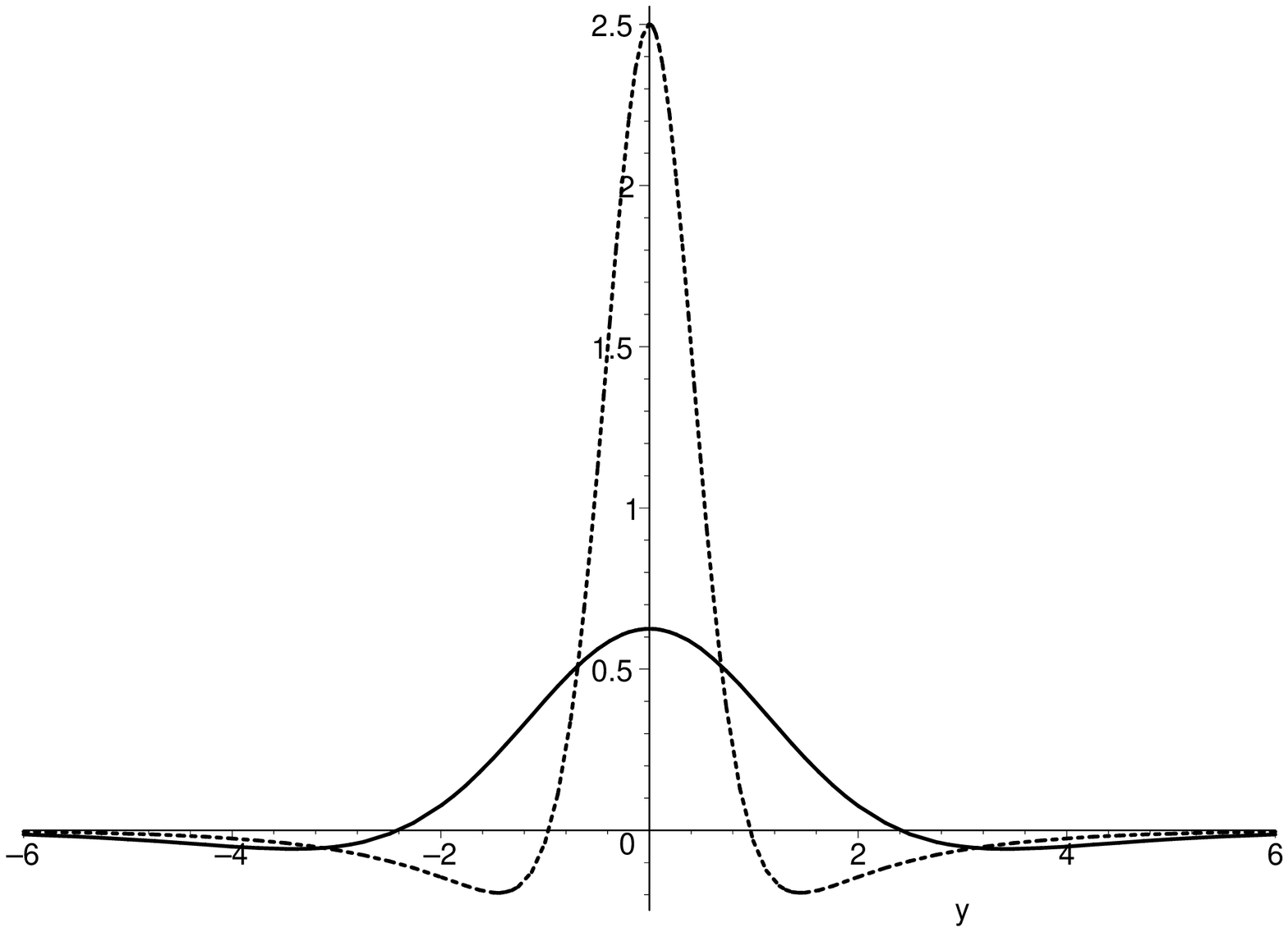}
\vspace{0.3cm}
\caption{Warp factor (left panel) for $a=1$, and
$b=1/2$ (solid line), $b=1$ (dashed line), and the corresponding
energy densities (right panel) for the scalar fields in curved
space-time for the model described by Eq.~(2.15).}\label{fig_wphi4}
\end{figure}

Another example is the well-known $\lambda\phi^4$ model obtained
with \be\label{wphi4} W=2ab(\phi-b^2\phi^3/3), \ee which gives the
scalar potential \be V(\phi)=\frac12 a^2b^2(1-b^2\phi^2)^2-\frac43
a^2b^2\phi^2(1-\frac{ b^2}{3}\phi^2)^2. \ee For this model one finds
\be \phi(y)=\pm\frac{1}{b}\tanh(ab^2y), \ee and \be\label{ay2}
A(y)=\frac{4}{9b^2}\ln\left({\rm
sech}(ab^2y)\right)-\frac{1}{9b^2}\tanh^2(ab^2y). \ee In examples
above $a,b,$ are constants. Note that these two examples although
they represent branes with finite energy, since their energy density
are localized (Figs.~\ref{fig_wsech} and \ref{fig_wphi4},
respectively) the kink profile behaves completely different at
asymptotic limits. In the first example the kink connects vacua at
infinity, whereas the well-known $\lambda\phi^4$ model connects the
two vacua $\phi_{vac}=\pm 1/b$. In any case the vacua are
supersymmetric since $W_\phi=0$ on them.

\subsection{Bent branes}

We now consider the general case of four dimensional AdS, or dS
geometry. Here the Einstein's equation give \bes\label{ee} \ben
A^{\prime\prime}+\Lambda e^{-2A}=-\frac23\phi^{\prime2}\label{ee2},
\\
A^{\prime2}-\Lambda
e^{-2A}=\frac16\phi^{\prime2}-\frac13V(\phi)\label{ee1}, \een \ees
for dS geometry ($\Lambda>0$) or AdS geometry ($\Lambda<0$). The
case of Minkowski spacetime is obtained in the limit $\Lambda\to0,$
which leads us back to Eqs.~(\ref{ee0}).

\begin{figure}[!ht]
\vspace{.3cm}
\includegraphics[{height=7.5cm,width=7cm,angle=-90}]{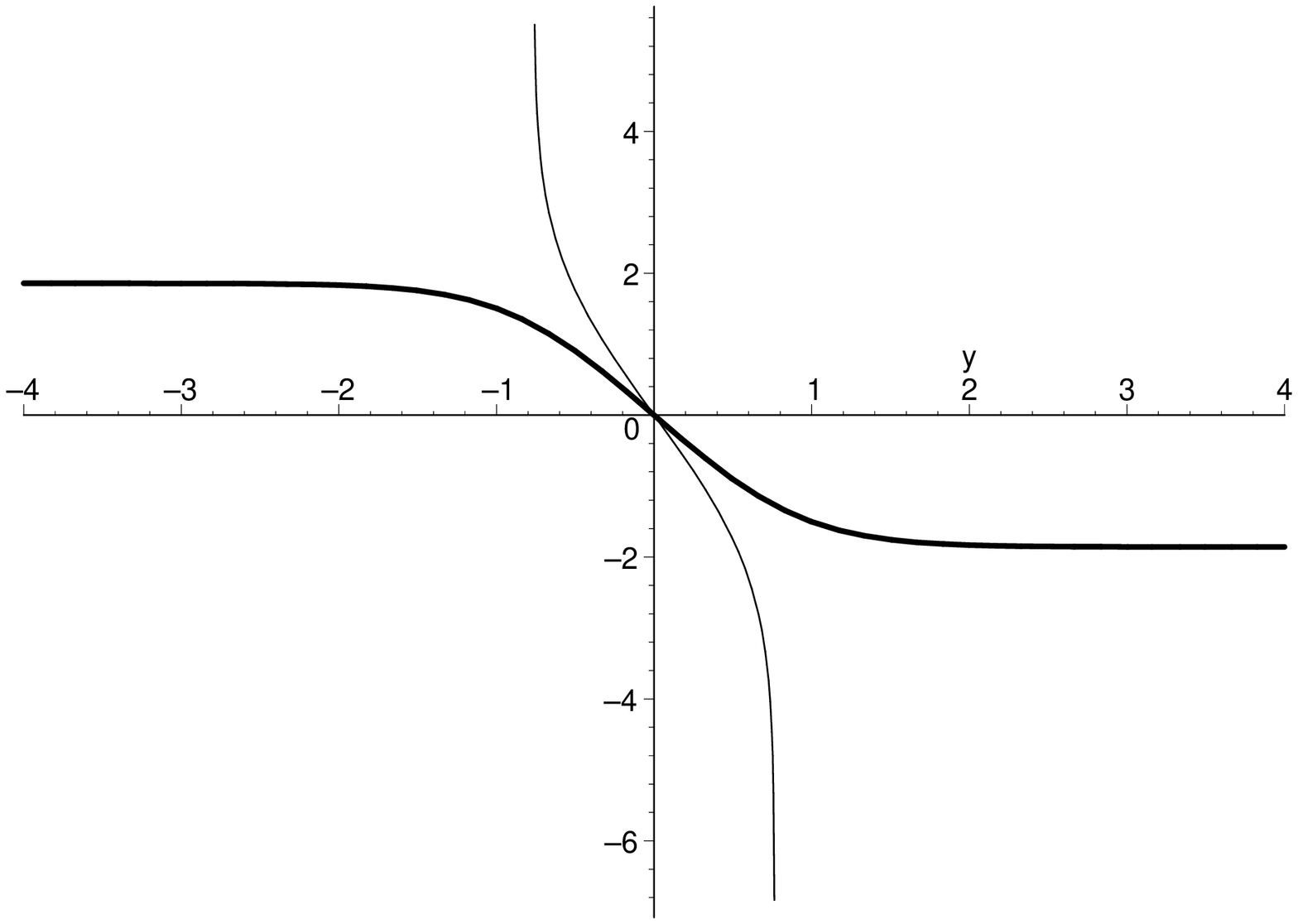}
\includegraphics[{height=7.5cm,width=7cm,angle=-90}]{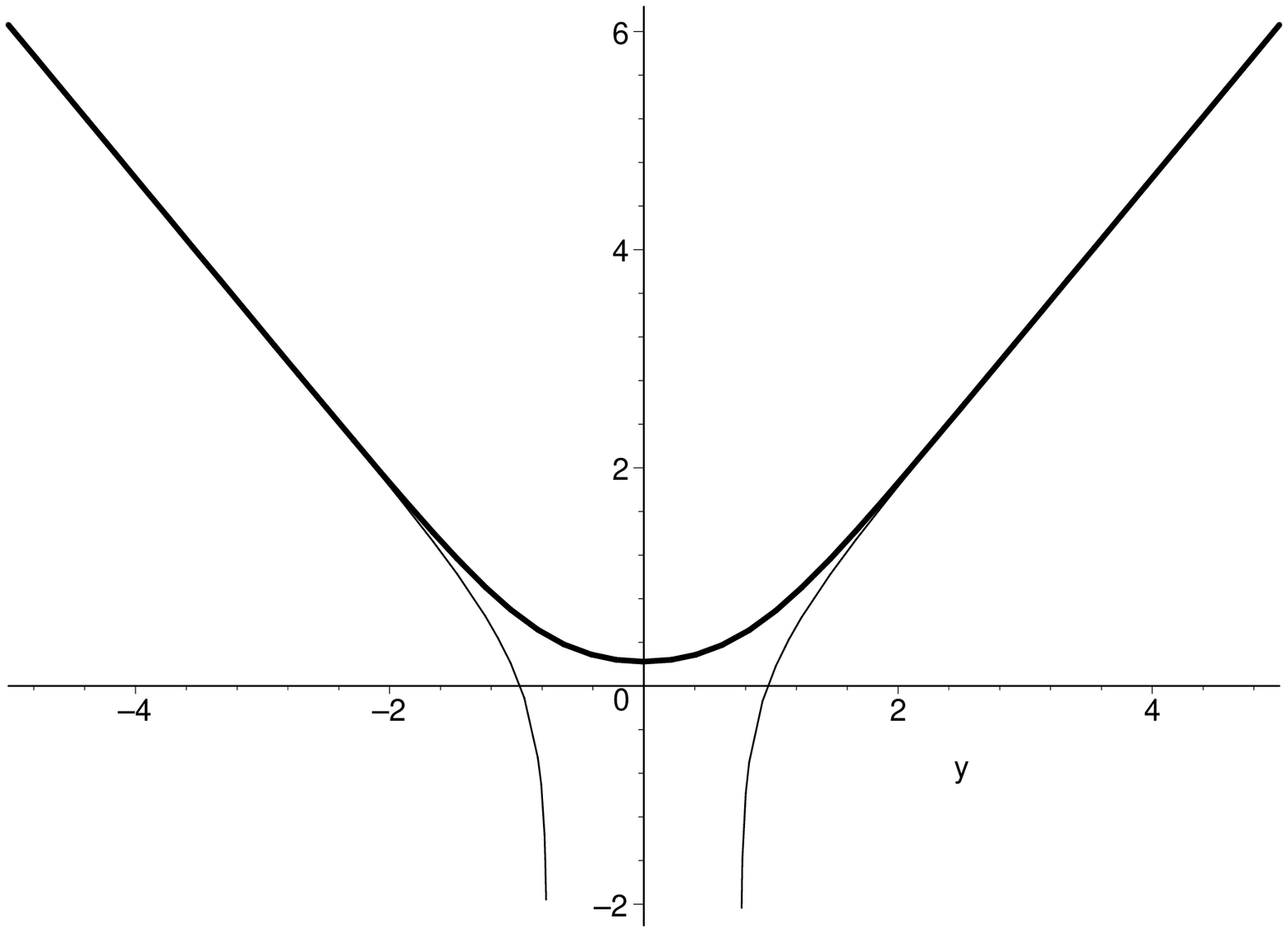}
\vspace{0.3cm} \caption{The kink profile (left panel) and $A(y)$
(right panel) are singular (thin line) for $\Lambda_{dS}>ab/\alpha$
and non-singular for $\Lambda_{AdS}<-ab/\alpha$ (thick line) where
$a=1,b=2/\sqrt{3},\,\alpha=1$.}\label{fig5c}
\end{figure}

The presence of $\Lambda$ makes the problem much harder. Interesting
investigations have been already appeared in Refs.~{\cite{fn,celi}}
and in references therein. Here, however, we follow another route.
The key issue springs signalizing for the need of a new constraint,
and we suggest that \be\label{A2} A^\prime=-\left(\frac13
W+\frac13\Lambda\gamma Z\right), \ee \be\label{phi2}
\phi^\prime=\frac12(W_\phi+\Lambda(\alpha+\gamma) Z_\phi), \ee where
$Z=Z(\phi)$ is a new and in principle arbitrary function of the
scalar field, to respond for the presence of the cosmological
constant, and $\alpha,\gamma$ are constants. In this case the
potential is given by \be V=\frac18(W_\phi+\Lambda(\alpha+\gamma)
Z_\phi)(W_\phi+\Lambda(\gamma-3\alpha)
Z_\phi)-\frac13(W+\Lambda\gamma Z)^2, \ee and we have to include the
constraint \be\label{cons} W_{\phi\phi}Z_\phi+W_\phi
Z_{\phi\phi}+2\Lambda(\alpha+\gamma)Z_\phi Z_{\phi\phi}-\frac43
Z_\phi(W+\Lambda\gamma Z)=0, \ee to obtain \be
A(y)=-\frac12\ln\left(\mp\frac{\alpha}{6}\left(W_{\phi}Z_{\phi}+
\Lambda(\alpha+\gamma)Z^2_{\phi}\right)\right), \ee which opens
diverse possibilities to obtain first-order formalism for
braneworlds with nonzero cosmological constant. To illustrate this
result, let us first consider the case $Z=\phi$, and
\be\label{wsinh1} W=a\sinh(b\phi)-\Lambda\gamma\phi, \ee where, from
Eq.~(\ref{cons}), $b=\pm2/\sqrt{3}$, and we have the potential
\be\label{Vsinh1}
V=\frac18\left(ab\cosh(b\phi)+\Lambda\alpha\right)\left(ab\cosh(b\phi)-3\Lambda\alpha\right)
-\frac13 a^2\sinh^2(b\phi). \ee In this case, the problem is solved
with \be\label{kink} \phi(y)=\pm\frac2b{\rm
arctanh}\left(\sqrt{\frac{ab+\Lambda\alpha}{ab-\Lambda\alpha}}\tan\left(\frac14
b\sqrt{a^2b^2-\Lambda^2\alpha^2}\;y\right)\right), \ee
\be\label{aysinh1}
A(y)=-\frac12\ln\left(\frac{\alpha}{6}\frac{a^2b^2-\Lambda^2\alpha^2}
{ab+\Lambda\alpha-2ab\,{\cos\,}^2\left(\frac14\;
b\sqrt{a^2b^2-\Lambda^2\alpha^2}\;y\right)}\right). \ee The kink
profile (\ref{kink}) and the brane geometry (\ref{aysinh1}) have the
main properties depicted in Figs.~\ref{fig5c}. For branes with
$AdS_4$ geometry, i.e., $\Lambda<-ab/\alpha$ the kink is smooth,
whereas for branes with $dS_4$ geometry, i.e., $\Lambda>ab/\alpha$,
the kink becomes `singular' in the sense that it diverges around
$y^*\!=\!\pm4{\rm arctanh}[(ab-\Lambda\alpha)/
\sqrt{-a^2b^2+\Lambda^2\alpha^2}]/(b\sqrt{-a^2b^2+\Lambda^2\alpha^2})$.
See also, e.g., Ref.~\cite{wang}, for another type of singularity in
$dS_4$ branes. We assume $ab>0,\,\alpha>0$. The case
$-ab/\alpha<\Lambda<ab/\alpha$, which necessarily includes branes
with Minkowski geometry, i.e., $\Lambda=0$, gives an array of
singular kinks. This is a nice example where a singular brane with
naked singularity (Fig.~\ref{fig5c}, $thin$ line at right panel) can
be smoothed out into another brane (Fig.~\ref{fig5c}, $thick$ line
at right panel) by turning on a negative cosmological constant on
the brane.

Asymptotically the scalar field $\phi$ describing the $smooth$ kink
goes to the vacuum sector $\phi_{vac}$ ($finite$ constant) and the
scalar potential approaches the five-dimensional cosmological
constant\be
V(\phi_{vac})\equiv\Lambda_5=-\frac{1}{3}(W(\phi_{vac})+\Lambda\gamma
Z(\phi_{vac}))^2=-3A'(\pm
\infty)^2=\frac{a^2}{3}-\frac{\Lambda^2\alpha^2}{4},\ee where we
have used $b=2/\sqrt{3}$, the potential (\ref{Vsinh1}), and the
solution (\ref{kink})-(\ref{aysinh1}) at asymptotic limits. Note
that these are supersymmetric vacua since they satisfy
${\phi'}_{vac}=\pm(1/2)(W_\phi+\Lambda(\alpha+\gamma)Z_\phi)|_{vac}=0$
(Recall that for flat branes, i.e., $\Lambda=0$, the supersymmetric
vacua satisfy the simple condition $W_\phi=0$.) Furthermore they
correspond to an asymptotic $AdS_5$ geometry, i.e., $\Lambda_5<0$,
because $\alpha^2\Lambda^2>a^2b^2$.

\noindent We end this section by considering another model obtained
with $Z=W$ and superpotential \be\label{wsinh2} W=a\sinh(b\phi),\ee
where, from Eq.~(\ref{cons}), we have
$b=\pm\sqrt{6(1-\Lambda\alpha)(1+\Lambda(\alpha+\gamma))}/3(1+\Lambda(\alpha+\gamma))$.
Here, we find the following scalar potential \be
V=\frac1{12}(1-\Lambda\alpha)(1+\Lambda(\gamma-3\alpha)a^2\cosh^2(b\phi)-\frac13
a^2(1+\Lambda\gamma)^2\sinh^2(b\phi). \ee In this case, the problem
is solved with \be\label{kink2} \phi(y)=\pm\frac1b{\rm
arcsinh}\left(\tan\left(\frac13a(1-\Lambda\alpha)\;y\right)\right),
\ee \be\label{aysinh2} A(y)=-\frac12\ln\left(-\frac19\alpha
a^2(1-\Lambda\alpha)\;\sec^2\left(\frac13
a(1-\Lambda\alpha)\;y\right)\right). \ee Note that the periodic kink
(\ref{kink2}) is singular and the metric has naked singularity at
$y^{*}=\pm3\pi/2a(1-\Lambda\alpha)$, for any cosmological constant
$\Lambda$. For $\alpha>0$, we only have $dS_4$ branes, since only
positive cosmological constant are allowed in this case, and for
$\alpha<0$ we may have $AdS_4$ or $dS_4$ branes, because the
cosmological constant on the brane can assume the values
$-1/|\alpha|<\Lambda<0$ or $\Lambda>0$, respectively. Similar models
have been first introduced in Refs.~\cite{g1,g2,abl2006}.

\section{Two scalar fields}
\label{two}

We now extend the above procedure to the case of two or more real
scalar fields with standard dynamics. We first investigate the
important case where the brane has 4d Minkowski geometry and later
we extend the analysis to 4d AdS and dS geometries. Here we have to
change the Lagrangian density to the form \be {\cal
L}=\frac12\partial_\mu\phi\partial^\mu\phi+
\frac12\partial_\mu\chi\partial^\mu\chi+
\frac12\partial_\mu\rho\partial^\mu\rho+...
+\frac12\partial_\mu\zeta\partial^\mu\zeta-V(\phi,\chi,\rho,...,\zeta).
\ee

\subsection{Flat branes}

For flat brane geometry in many fields theory we get the new set of
equations \bes\label{em2f} \ben
&&\phi^{\prime\prime}+4A^{\prime}\phi^{\prime}
=V_\phi\;,\;\;\;\chi^{\prime\prime}+4A^{\prime}\chi^{\prime}
=V_\chi\;,\;\;\;\rho^{\prime\prime}+4A^{\prime}\rho^{\prime}
=V_\rho\;,\;...,\;\;\;\zeta^{\prime\prime}+4A^{\prime}\zeta^{\prime}=V_\zeta,
\\
&&A^{\prime\prime}=-\frac23\phi^{\prime2}-\frac23\chi^{\prime2}
-\frac23\rho^{\prime2}-\;...-\frac23\zeta^{\prime2},
\\
&&A^{\prime2}=\frac16\phi^{\prime2}+\frac16\chi^{\prime2}+\frac16\rho^{\prime2}
+\;...+\frac16\zeta^{\prime2}-\frac13V. \een\ees As before, we
insist with $A^\prime=-W/3,$ but now $W=W(\phi,\chi,\rho,...,\zeta)$
suggests that we write the two first-order equations \be
\phi^\prime=\frac12W_\phi\;,\;\;\;\;\;\chi^\prime=
\frac12W_\chi\;,\;\;\;\;\;\rho^\prime=\frac12
W_\rho\;,\;...,\;\;\;\;\;\zeta^\prime=\frac12W_\zeta, \ee and the
potential is now given by \be\label{p2} V=\frac18W^2_\phi+\frac18
W^2_\chi+\frac18 W^2_\rho+...+\frac18 W^2_\zeta-\frac13 W^2. \ee It
is not hard to show that solutions of the above first-order
equations also solve the set of Eqs.~(\ref{em2f}) for the potential
(\ref{p2}). The above procedure opens interesting possibilities for
setups of coupled fields and branes with flat geometry.

\begin{figure}[!ht]
\vspace{.3cm}
\includegraphics[{height=7.5cm,width=7cm,angle=-90}]{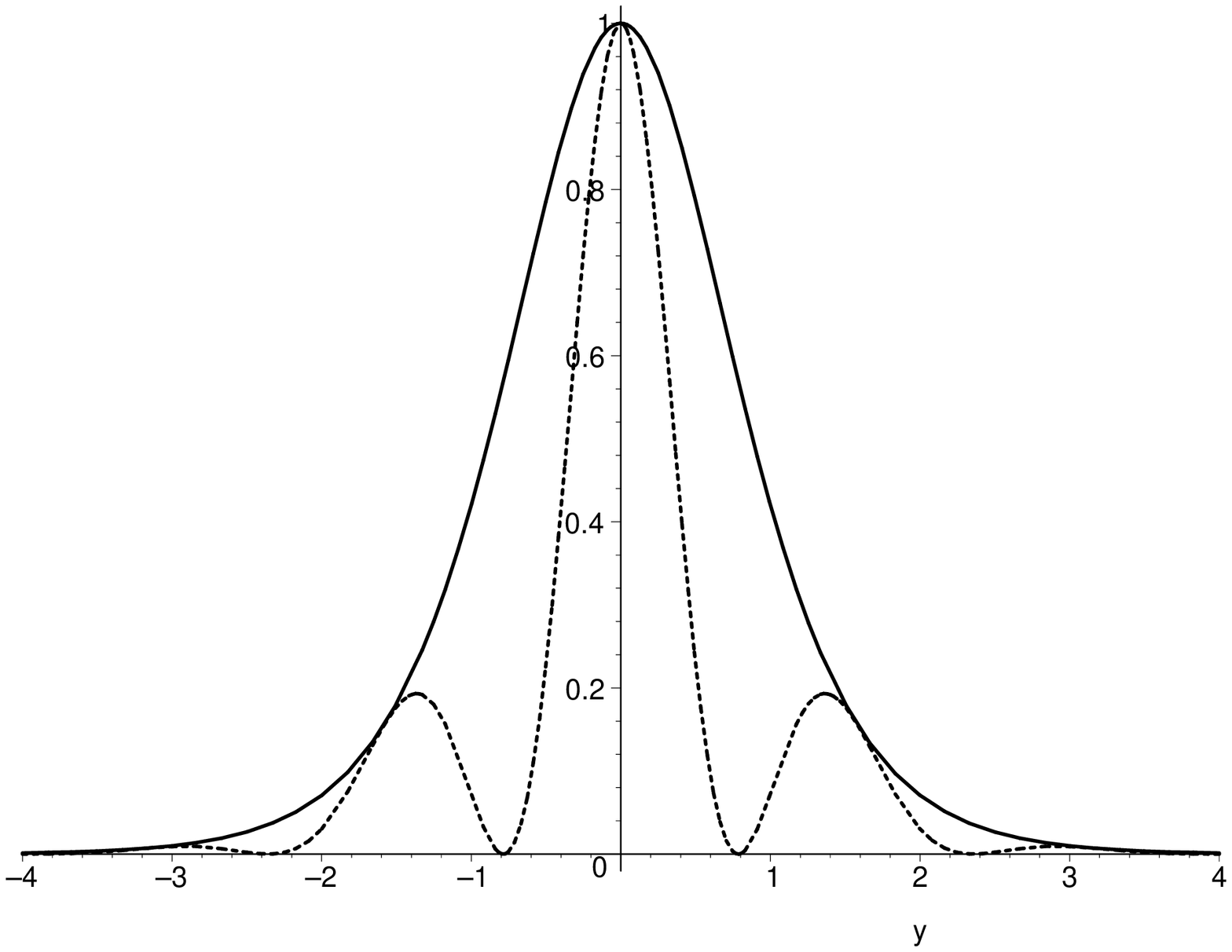}
\includegraphics[{height=7.5cm,width=7cm,angle=-90}]{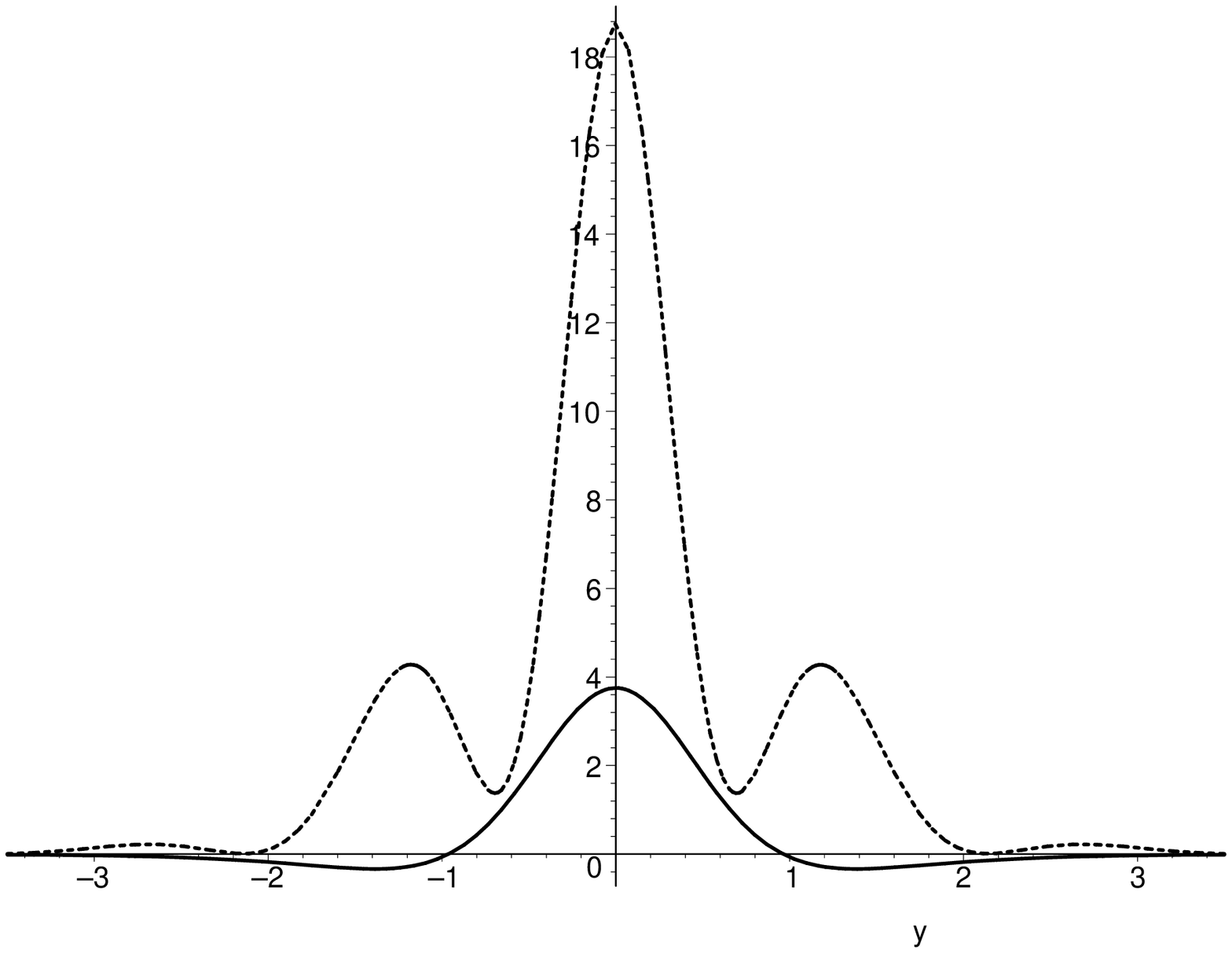}
\vspace{0.3cm} \caption{Warp factor (left panel) for $c=0,$ and $2$
(solid line, and dashed line) and the corresponding energy densities
(right panel) for the scalar fields in curved space-time for the
model described by Eq.~(3.5) with $b=d=\sqrt{2/3}$ and
$a=p=q=1$.}\label{fig6}
\end{figure}

In the case of two scalar fields, for instance, if we consider an
additive $W,$ that is, if we take
$W(\phi,\chi)=W_1(\phi)+W_2(\chi),$ we get the potential in the form
$V(\phi,\chi)=V_1(\phi)+V_2(\chi)-(2/3)W_1(\phi)W_2(\chi).$ It shows
that the interactions appear as the product of the two independent
$W_1$ and $W_2.$ As an example we consider \be\label{wsinsinh}
W=3a\sin(b\phi)+3c\sinh(d\chi), \ee where $a,b,c,d$ are constants.
The scalar potential for such superpotential is \be V=\frac98
a^2b^2\cos^2(b\phi)+\frac98
c^2d^2\cosh^2(d\chi)-3\left(a\sin(b\phi)+c\sinh(d\chi)\right)^2, \ee
such that the solution of the decoupled first order equations reads
\bes \ben \phi(y)=\pm\frac1b{\rm arcsin}[\tanh(\frac32 ab^2y)],
\\
\chi(y)=\pm\frac1d{\rm arcsinh}[\tan(\frac32 cd^2y)], \een \ees and
\be\label{ay4} A(y)=-\frac{2}{3b^2}\ln[q\cosh(\frac32
ab^2y)]-\frac{2}{3d^2}\ln[p\,\sec(\frac32 cd^2y)], \ee where $p$ and
$q$ are real positive constants. In this case, the metric has naked
singularity at $y^*=\pm\pi/3cd^2$. The case where $d=0$ or $b=0$
reduces to the case of one scalar field system, as it has been
previously found in Refs.~\cite{g1,g2} and \cite{abl2006},
respectively. The warp factor $e^{2A}$ and energy density for this
solution are depicted in Fig.~\ref{fig6}. Note that the case $c=0$
is equivalent to have only the field $\phi$, and there is no naked
singularity on the metric. By considering only the scalar field
$\phi$ component, i.e., $c=0$, the vacuum is achieved in the
asymptotic limits $\phi_{vac}=\phi(\pm\infty)$. The metric
asymptotically describes an $AdS_5$ space whose cosmological
constant is $\Lambda_5\equiv V(\phi_{vac})=-3a^2$. On the other
hand, by turning on both scalar field components, i.e.,
$a,b,c,d\neq0$, the solution becomes singular and periodic.

\begin{figure}[!ht]
\vspace{.3cm}
\includegraphics[{height=7.5cm,width=6cm,angle=-90}]{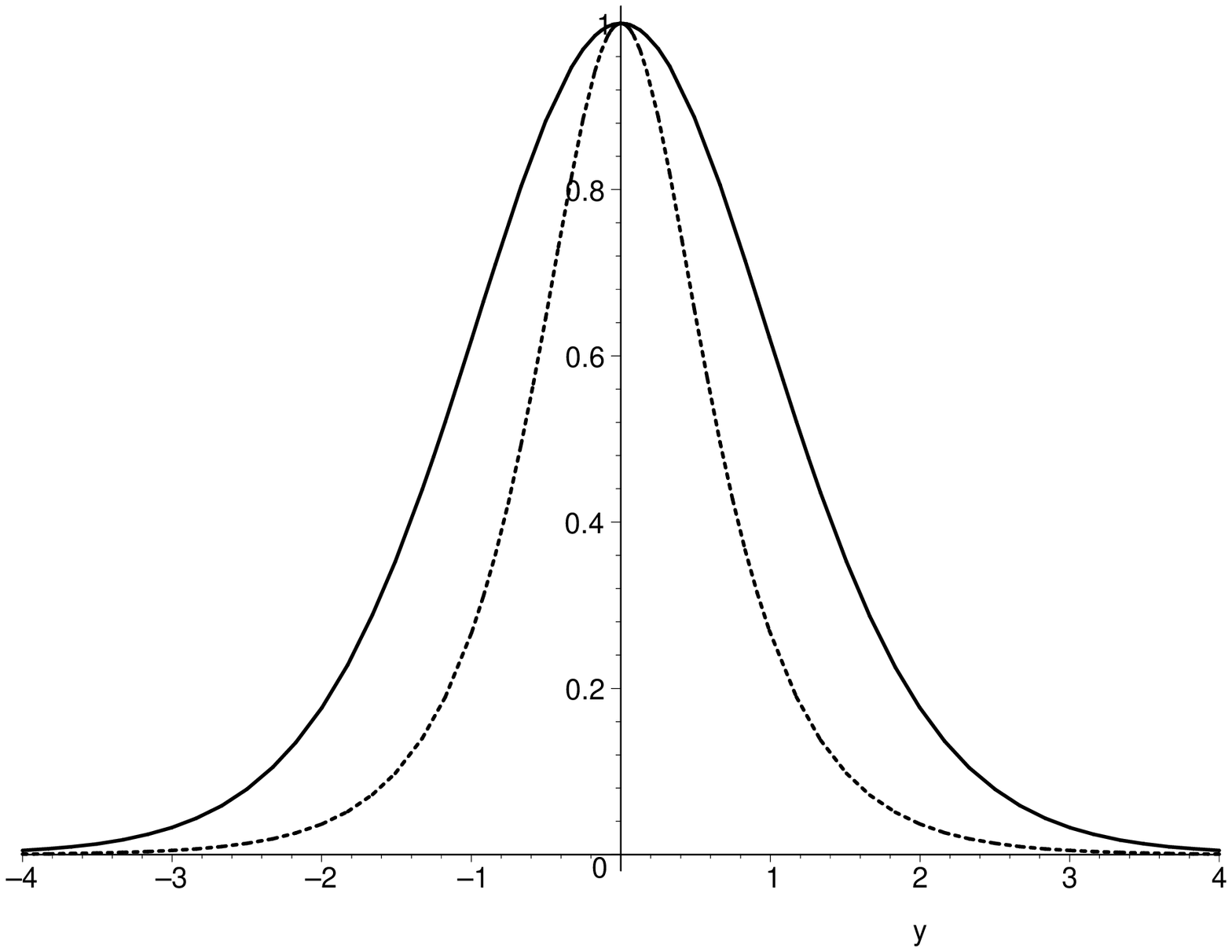}
\includegraphics[{height=7.5cm,width=6cm,angle=-90}]{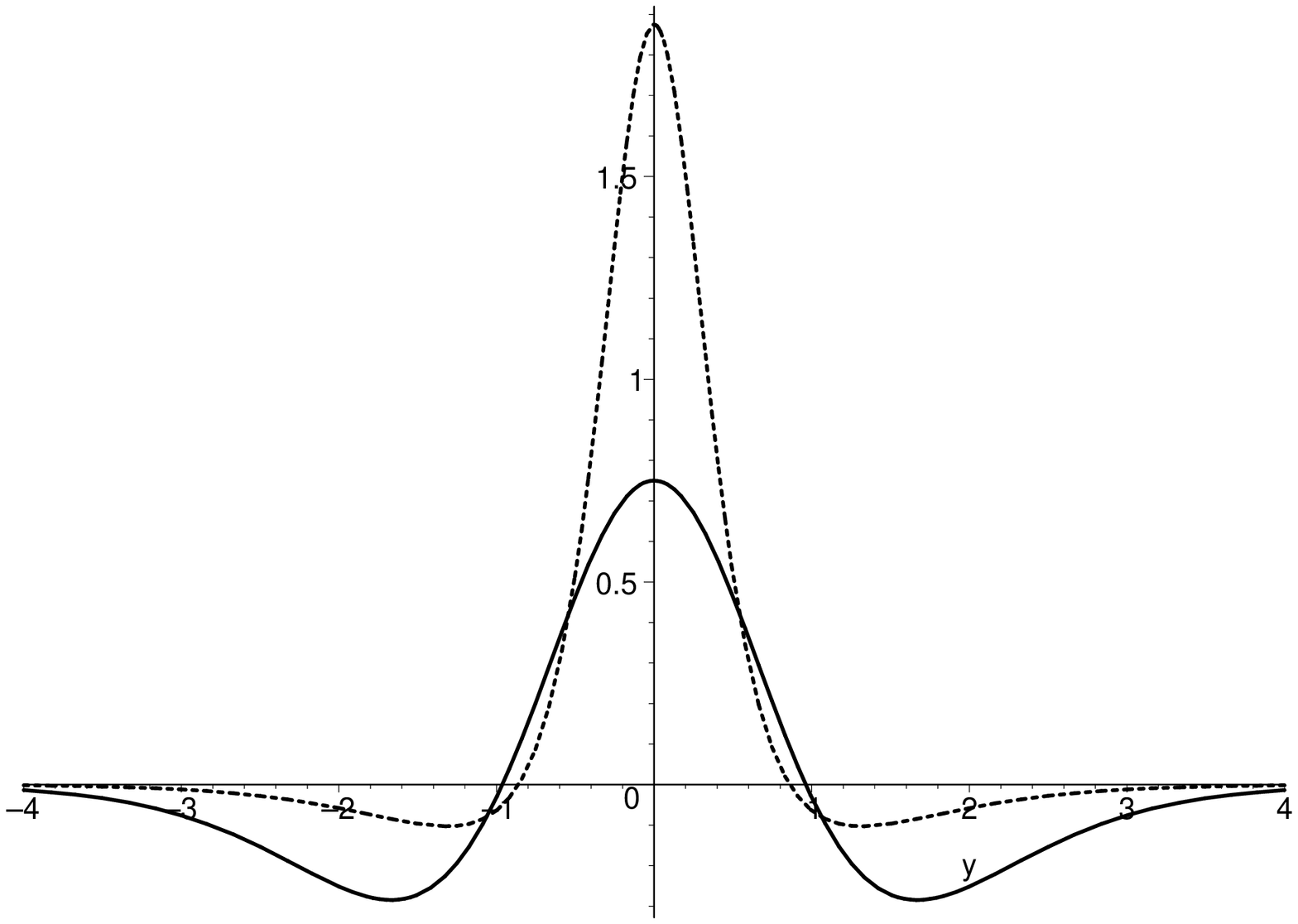}
\vspace{0.3cm} \caption{Warp factor (left panel) with $a=1$ and
$b=1/\sqrt{3}$ (solid line), $b=2/\sqrt{3}$ (dashed line), and the
corresponding energy density (right panel)}\label{fig7}
\end{figure}

Another possibility, that couples the scalar fields in the
superpotential, is given by \be\label{wsincos}
W=3a\sin(b\phi)\cos(b\chi). \ee In this case we have \be V=\frac98
a^2b^2\left(\cos^2(b\phi)\cos^2(b\chi)+\sin^2(b\phi)\sin^2(b\chi)\right)-3
a^2\sin^2(b\phi)\cos^2(b\chi). \ee We can find the solution of the
coupled first order equations by using the orbits
$\cos(b\phi)=C\;\sin(b\chi)$ being $C$ a real constant. For $C=0$ we
have \be \phi(y)=(2m+1)\frac{\pi}{2b},\;\;\chi(y)=\pm\frac1b
\arccos\left(\tanh\left(\frac32
ab^2y\right)\right)+k\;\frac{\pi}{b}, \ee or \be
\phi(y)=\pm\frac1b\arcsin\left(\tanh\left(\frac32ab^2y\right)\right)+
k\;\frac{\pi}{b},\;\;\;\;\;\chi(y)=m\;\frac{\pi}{b}, \ee where $m$
and $k$ are integer, from that \be\label{ay5a}
A(y)=-\frac{2}{3b^2}\ln[\;q\cosh(\frac32 ab^2y)\;], \ee where the
constant $q>0$. The warp factor $e^{2A}$ and energy density for this
solution is shown in Fig.~\ref{fig7}. These brane solutions are
supersymmetric in the sense that asymptotically, i.e., at the vacuum
$\phi_{vac}=\phi(y=\pm\infty),\chi_{vac}=\chi(y=\pm\infty)$, we find
$W_\phi,W_\chi=0$. The bulk is asymptotically a five-dimensional
$AdS_5$ space-time whose cosmological constant is $\Lambda_5\equiv
V(\phi_{vac},\chi_{vac})=-3a^2$.

\begin{figure}[!ht]
\vspace{.3cm}
\includegraphics[{height=7.5cm,width=6cm,angle=-90}]{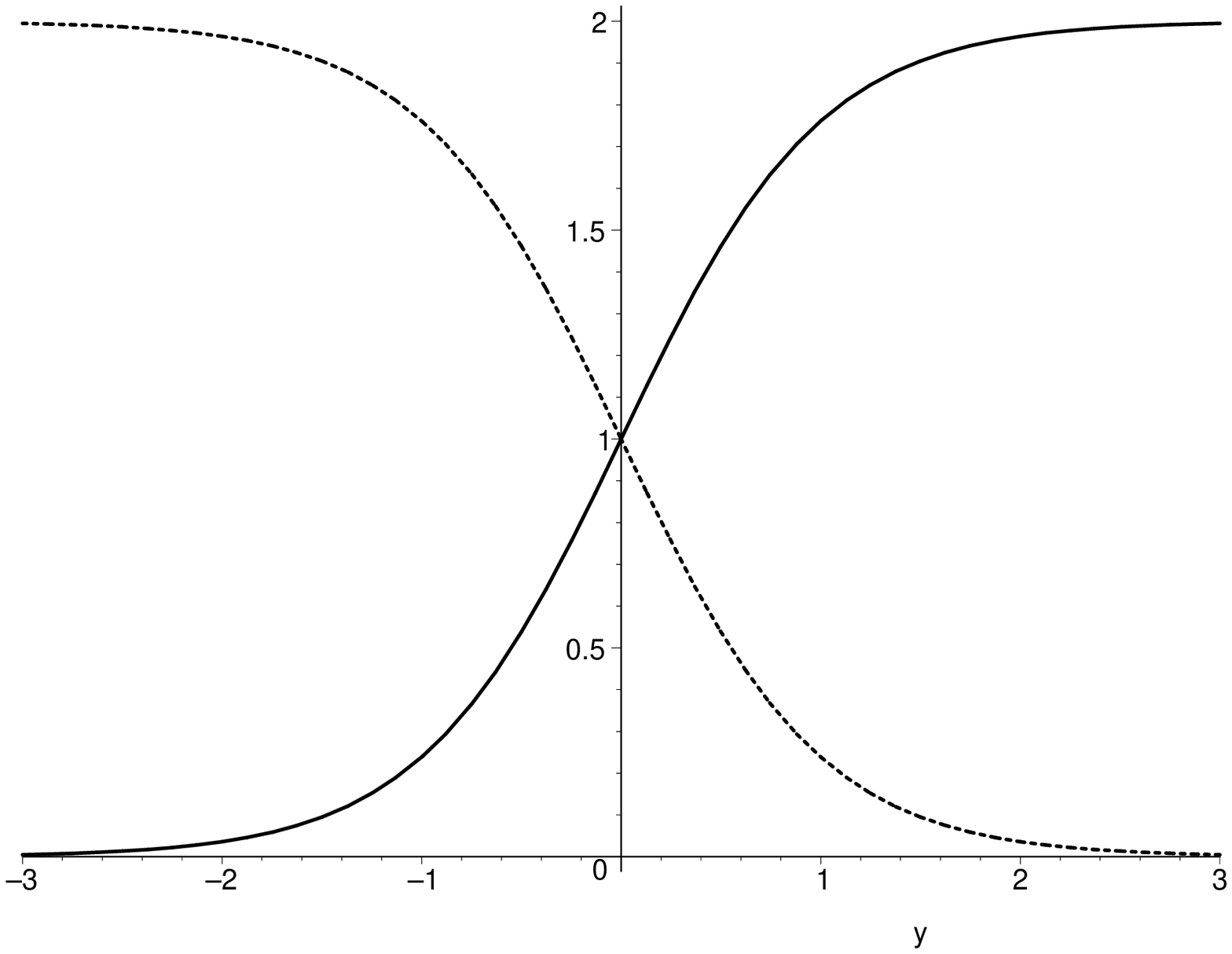}
\includegraphics[{height=7.5cm,width=6cm,angle=-90}]{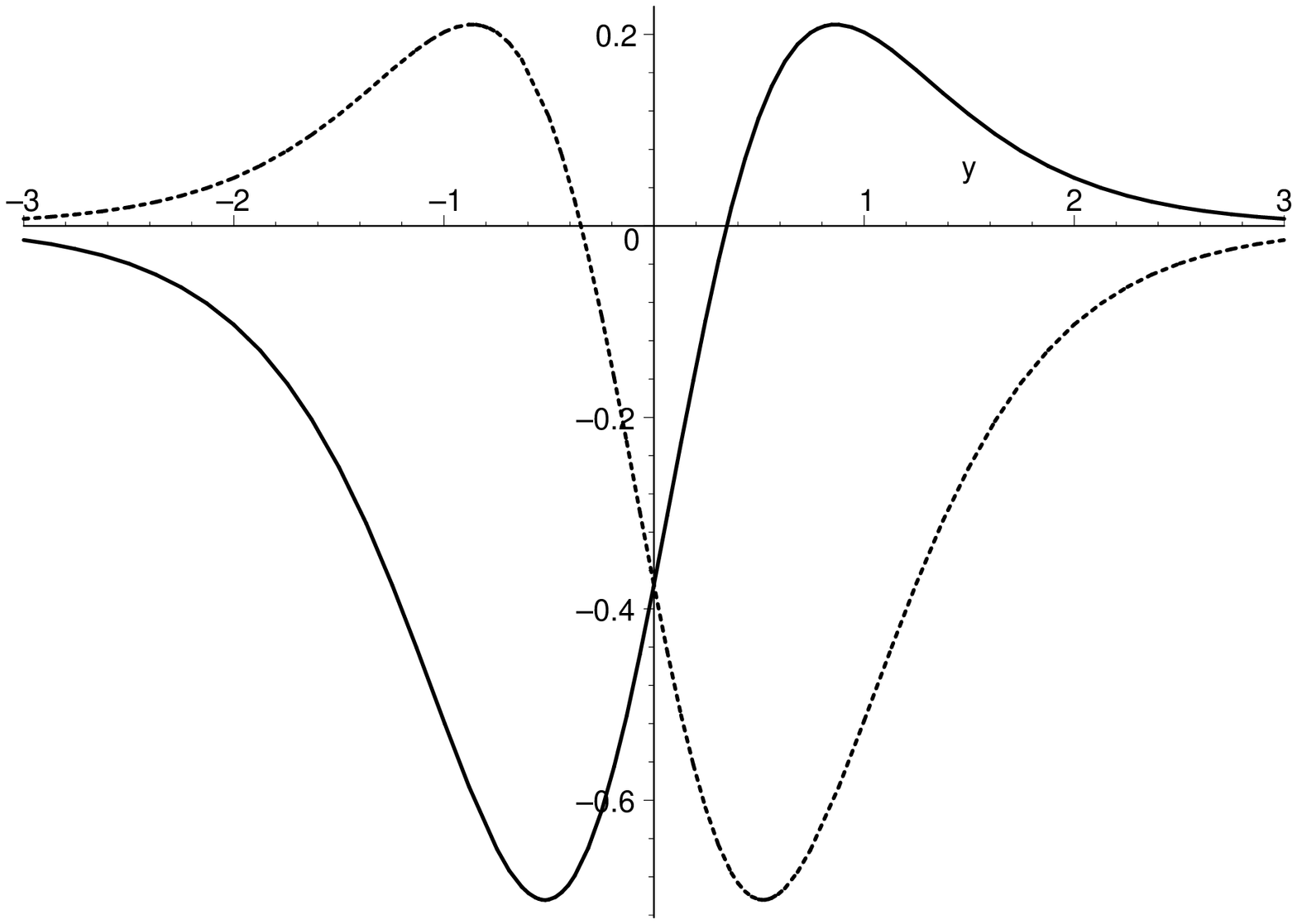}
\vspace{0.3cm} \caption{Warp factor of asymmetric BPS branes
connecting asymptotically $AdS_5-\mathbb{M}_5$ spaces (solid line)
and $\mathbb{M}_5-AdS_5$ spaces (dashed line), with $a=1$ and
$b=2/\sqrt{3}$ (left panel), and the corresponding energy densities
(right panel).}\label{fig8}
\end{figure}

\begin{figure}[!ht]
\vspace{.3cm}
\includegraphics[{height=7.5cm,width=6cm,angle=-90}]{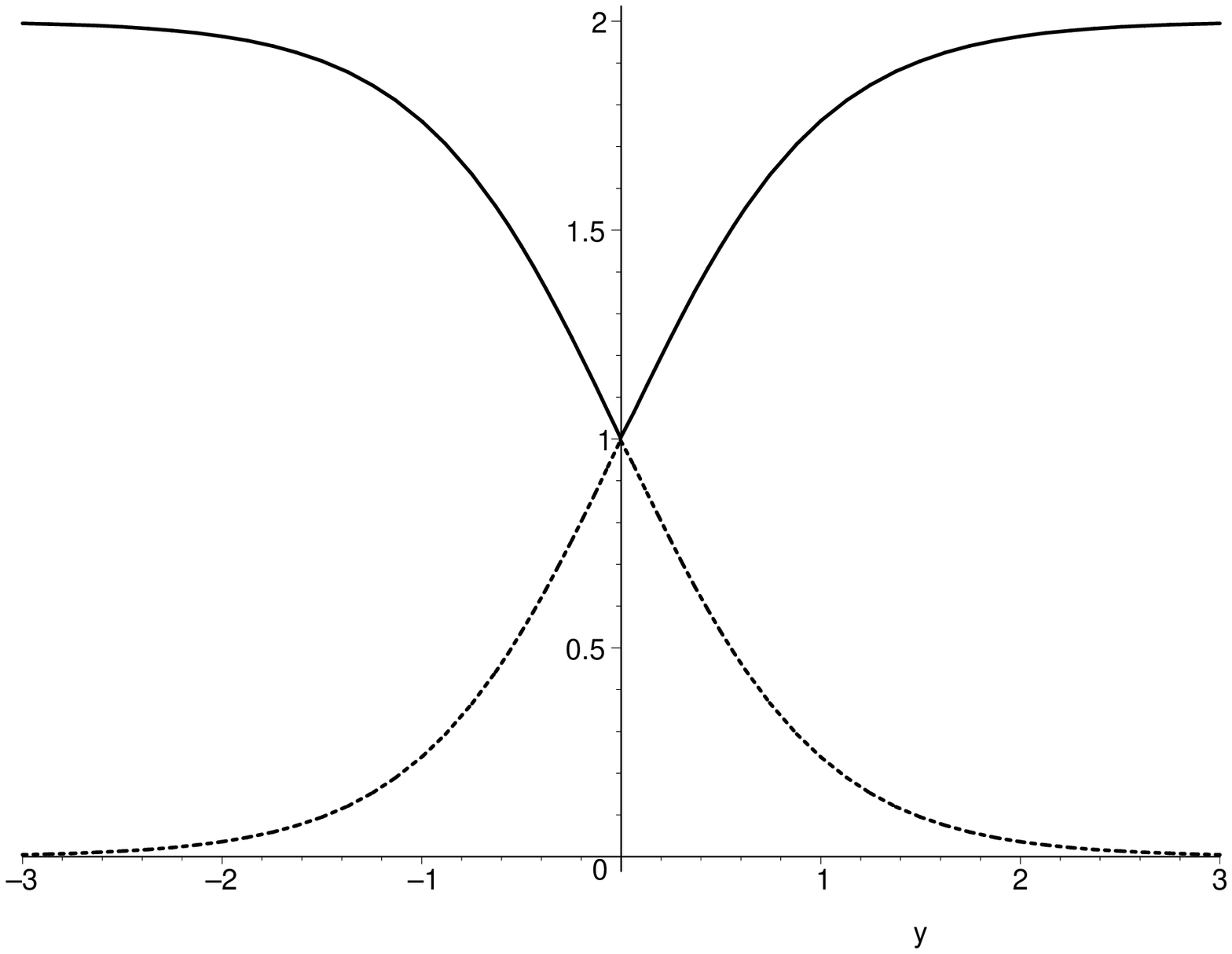}
\includegraphics[{height=7.5cm,width=6cm,angle=-90}]{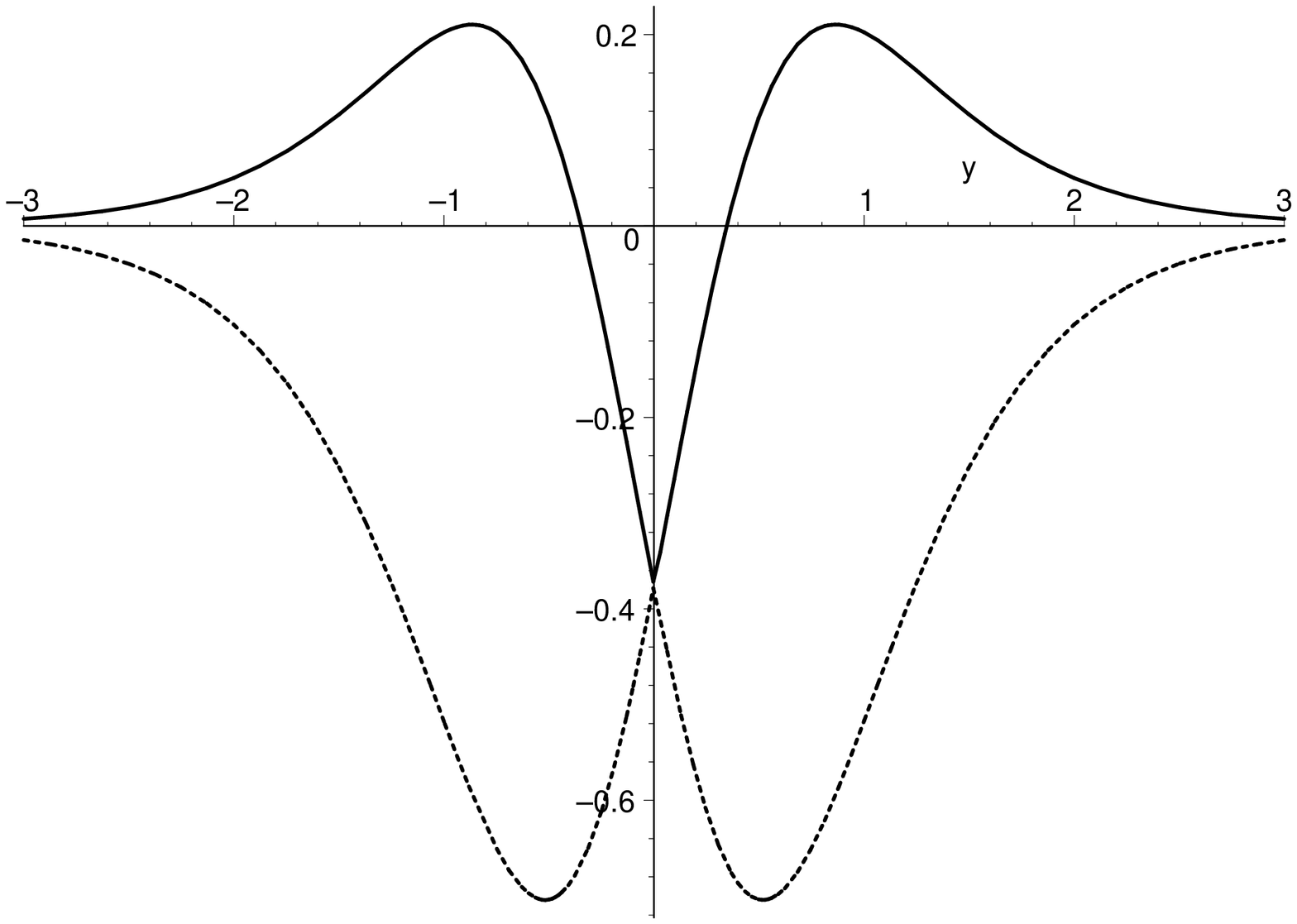}
\vspace{0.3cm} \caption{Warp factor of symmetric non-BPS branes
connecting asymptotically $\mathbb{M}_5-\mathbb{M}_5$ spaces (solid
line) and $AdS_5-AdS_5$ spaces (dashed line), with $a=1$ and
$b=2/\sqrt{3}$ (left panel), and the corresponding energy densities
(right panel).}\label{fig9}
\end{figure}

For $C=1$ we have \be\label{pq1}
\phi(y)=\pm\frac1{2b}\arccos\left(\tanh(\frac34 ab^2y)\right)+(k+1)
\;\frac{\pi}{2b},\;\;\;\;\chi(y)=\pm\frac1{2b}\arccos\left(\tanh(\frac32
ab^2y)\right)+k\;\frac{\pi}{2b}, \ee from that \be\label{ay5b}
A(y)=(-1)^{k+1}\frac{ay}{2}+\frac{2}{3b^2}\ln[\;q\;{\rm
sech}(\frac34ab^2y)\;], \ee where the constant $q>0$. Now we have
asymmetric (Fig.~\ref{fig8}) and symmetric (Fig.~\ref{fig9}) branes.
Let us now concern about the rich vacuum structure and geometry of
these solutions. The supersymmetric vacua satisfying $W_\phi=0$,
$W_\chi=0$ are connected by BPS and non-BPS branes. The
superpotential $W(\phi,\chi)$ evaluated at the vacua
$(\phi_{vac},\chi_{vac})$ gives the asymptotic behavior of the
geometry governed by equation $A'=-W/3$. At the vacua, i.e., for
$y=\pm\infty$, the kink solutions for $k$ odd or even give us (i)
$W_{odd}^+=0$, (ii) $W_{odd}^-=3a$, (iii) $W_{even}^+=3a$ and (iv)
$W_{even}^-=0$. At these supersymmetric vacua the 5d cosmological
constant
$\Lambda_5\!\equiv\!V(\phi_{vac},\chi_{vac})\!=\!-(1/3)[W^{\pm}_{even/odd}]^{\,2}\leq0$
asymptotically characterizes five-dimensional Minkowski
($\mathbb{M}_5$) or anti-de Sitter ($AdS_5$) spaces. The cases
(i)-(ii) and (iii)-(iv) describe $asymmetric$ branes connecting
asymptotically $AdS_5-\mathbb{M}_5$ spaces and $\mathbb{M}_5-AdS_5$
spaces, respectively
--- See Fig.~\ref{fig8}. For further discussions on asymmetric branes
see, e.g., Refs.~\cite{melfo1,melfo2,ts,pad,gort}. On the other
hand, we can patch together $even$ and $odd$ solutions to form
$\mathbb{Z}_2$ symmetric branes. The cases (ii)-(iii) and (i)-(iv)
describe such $symmetric$ branes connecting asymptotically
$AdS_5-AdS_5$ spaces and $\mathbb{M}_5-\mathbb{M}_5$ spaces,
respectively --- See Fig.~\ref{fig9}. These symmetric branes are
clearly non-BPS branes, because the BPS bound
\cite{cvetic92,cvetic93,cvetic97,bsr,bbet} $\sigma_{BPS}=|\Delta
W|=|W^{\pm}_{even/odd}-W^{\mp}_{odd/even}|=0$, in contrast with the
asymmetric  BPS branes whose BPS bound $\sigma_{BPS}=3a$. Locally,
the geometry of the branes around $y=0$ behaves according to the
following branches $A(y)\!\simeq\!(-1)^{k+1}ay/2$. We can patch
together two local branches along with the symmetric branes
$A(y)\!\simeq\!-a|y|/2$ (see warp factor in Fig.~\ref{fig9}
--- dashed line) and $A(y)\!\simeq\!a|y|/2$ (see warp factor in
Fig.~\ref{fig9} --- solid line).

\subsection{Bent branes}

We now consider the general case of branes with four dimensional
anti-de Sitter ($AdS_4$) or de Sitter ($dS_4$) geometry.  Let us
restrict ourselves to a two scalar field theory on the background
(\ref{metric}). The Einstein's equations give \bes\label{em} \ben
&&A^{\prime\prime}+\Lambda
e^{-2A}=-\frac23\phi^{\prime2}-\frac23\chi^{\prime2},\label{em2}
\\
&&A^{\prime2}-\Lambda
e^{-2A}=\frac16\phi^{\prime2}+\frac16\chi^{\prime2}-\frac13V(\phi,\chi),\label{em1}
\een \ees for $dS_4$ ($\Lambda>0$) or $AdS_4$ ($\Lambda<0$)
geometry. The case of Minkowski space-time is obtained in the limit
$\Lambda\to0,$ which leads us back to Eqs.~(\ref{em2f}).

The presence of the four dimensional cosmological constant $\Lambda$
makes the problem much harder, but it can be done following the same
way employed for one scalar field. In this sense, we suggest that
the problem of integrating second-order equation of motion can be
reduced to the set of first-order equations \be\label{sugest1}
A^\prime=-\frac13(W+\Lambda\gamma Z), \ee \be\label{sugest2}
\phi^\prime=\frac12(W_\phi+\Lambda(\alpha+\gamma)\; Z_\phi), \ee
\be\label{sugest3}
\chi^\prime=\frac12(W_\chi+\Lambda(\beta+\gamma)\; Z_\chi), \ee
where $Z=Z(\phi,\chi)$ is a new and in principle arbitrary function
of the scalar field, to respond for the presence of the cosmological
constant, and $\alpha,\beta,\gamma$ are constants. The suggestion
(\ref{sugest1})-(\ref{sugest3}) is consistent with
Eqs.~(\ref{em2})-(\ref{em1}) if the scalar potential is given by
\ben V(\phi,\chi)&=&\frac18(W_\phi+\Lambda(\alpha+\gamma)
Z_\phi)(W_\phi+\Lambda(\gamma-3\alpha) Z_\phi)+\nn &
&\frac18(W_\chi+\Lambda(\beta+\gamma)
Z_\chi)(W_\chi+\Lambda(\gamma-3\beta)
Z_\chi)-\frac13(W+\Lambda\gamma Z)^2, \een and if we impose the
following constraints \bes \ben & &\alpha W_\phi Z_{\phi\phi}+\alpha
Z_\phi W_{\phi\phi}+2\Lambda\alpha(\alpha+\gamma)Z_\phi
Z_{\phi\phi}+\frac12(\alpha+\beta)W_\chi Z_{\phi\chi}+\nn & &\beta
Z_\chi W_{\phi\chi}+\frac12 \Lambda(\beta+\gamma)(\alpha+3\beta)
Z_\chi Z_{\phi\chi}-\frac43\alpha Z_\phi(W+\Lambda\gamma Z)=0,
\\ \nn
& &\beta W_\chi Z_{\chi\chi}+\beta Z_\chi
W_{\chi\chi}+2\Lambda\beta(\beta+\gamma)Z_\chi Z_{\chi\chi}
+\frac12(\alpha+\beta)W_\phi Z_{\phi\chi}+\nn & &\alpha Z_\phi
W_{\phi\chi}+\frac12 \Lambda(\alpha+\gamma)(3\alpha+\beta) Z_\phi
Z_{\phi\chi}-\frac43\beta Z_\chi(W+\Lambda\gamma Z)=0. \een \ees
After such considerations we obtain \be
A(y)=-\frac12\ln\left(-\frac{\alpha}{6}\left(W_{\phi}Z_{\phi}
+\Lambda(\alpha+\gamma)Z^2_{\phi}\right)-\frac{\beta}{6}\left(W_{\chi}Z_{\chi}
+\Lambda(\beta+\gamma)Z^2_{\chi}\right)\right). \ee To illustrate
this new result, let us consider the case $Z=W$, with $\gamma=0$ and
$\beta=\alpha$. We take \be\label{wsin} W=3a\sin(b\phi+c\chi), \ee
where $a$, $b$, and $c$ are real constants with
$b^2+c^2=-2/3(1+\Lambda\alpha)$, and the scalar potential reads \be
V(\phi,\chi)=3a^2\left(\left(1-\frac14(1-3\Lambda\alpha)\right)
\cos^2(b\phi+c\chi)-1\right). \ee This potential have global and
local minima. As $(1-3\Lambda\alpha)<4$ there exist global minima
given by $b\phi+c\chi=\pm(2m+1)\pi/2$ and for
$(1-3\Lambda\alpha)>4$, there exist local minima at $b\phi+c\chi=\pm
m\pi$, where $m=0,1,2,3,...$. It is not difficult to notice that the
global minima are supersymmetric vacua because $W_\phi=Z_\phi=0$ and
$W_\chi=Z_\chi=0$ implies $b\phi_{vac}+c\chi_{vac}=\pm(2m+1)\pi/2$.
The first-order equations have solutions for the orbits
$b\;\chi=c\;\phi+C$, where $C$ is a real constant. For
$(1-3\Lambda\alpha)<4$ and $C=b/c\;n\pi$, or $(1-3\Lambda\alpha)>4$
and $C=b/c\;(2n+1)\pi/2$, we have the regular kinks \be
\phi(y)=\pm\frac1d{\rm
\arcsin}\left(\;\tanh(a\;y)\;\right)+k\frac{\pi}{d}, \ee and the
irregular kinks \be \phi(y)=\pm\frac1d{\rm
arcsin}\left(\;\coth(a\;y)\;\right)+k\frac{\pi}{d}. \ee Furthermore,
for $(1-3\Lambda\alpha)<4$ and $C=b/c\;(2n+1)\pi/2$, or
$(1-3\Lambda\alpha)>4$ and $C=b/c\;n\pi$, we have the regular kinks
\be \phi(y)=\pm\frac1d{\rm
arccos}\left(\;\tanh(a\;y)\;\right)+k\frac{\pi}{d}, \ee and the
irregular kinks \be \phi(y)=\pm\frac1d{\rm
arccos}\left(\;\coth(a\;y)\;\right)+k\frac{\pi}{d}, \ee where $k$
and $n$ are integer numbers and $d=-2/b(1+\Lambda\alpha)$. For
regular kink solutions we obtain \be\label{ay6}
A(y)=\ln\left[\sqrt{\frac{1}{\alpha}}\frac{1}{|a|}\cosh(a\,y)\right],
\ee with $\alpha>0$. Since $\alpha=(-\Lambda)^{-1}[2/3(b^2+c^2)+1]$,
$\Lambda<0$, that is, this solution represents a brane with $AdS_4$
geometry. On the other hand,  for irregular kink solutions we obtain
\be\label{ay7}
A(y)=\ln\left[\sqrt{\frac{1}{-\alpha}}\;\frac{1}{|a|}\;|\sinh(a\;y)|\right],
\ee with $\alpha<0$. Now, because
$\alpha=(-\Lambda)^{-1}[2/3(b^2+c^2)+1]$, $\Lambda>0$, this solution
represents a brane with $dS_4$ geometry.

It is instructive to notice that for two strongly coupled fields,
i.e., $b^2+c^2\gg1$
we find that $\alpha=(-\Lambda)^{-1}$. Furthermore, at the global
(or supersymmetric) vacua we find $W(\phi_{vac},\chi_{vac})=3a$ and
then the 5d cosmological constant is $\Lambda_5\equiv
V(\phi_{vac},\chi_{vac})=-(1/3)W^2=-3a^2=-3/L^2$, where we have
identified $a=1/L$, being $L$ the $AdS_5$ radius. Under such
considerations the solution (\ref{ay6}) reduces to the familiar
solution of a brane with $AdS_4$ geometry \cite{f,kr,cvetic93}
(here, however, there is no $\delta$-source for the brane):\ben
A(y)=\ln\left[\sqrt{-\Lambda}L\cosh\left(\frac{a\,y}
{L}\right)\right].\een Similarly, the solution of a brane with
$dS_4$ (\ref{ay7}) geometry can be written in the familiar form \ben
A(y)=\ln\left[\sqrt{\Lambda}L\sinh\left(\frac{a\,y}
{L}\right)\right].\een Although on one hand it is hard to find
explicit solutions for many scalar fields, on the other hand it is
straightforward to generalize the formalism above for $N$ scalar
fields. The scalar potential is given by
\ben\label{multiV}V(\phi_1,...,\phi_N)=\frac{1}{8}\sum_{i=1}^N\Big(\partial_i
W+\Lambda(\alpha_i+\gamma)\partial_iZ\Big)\Big(\partial_i
W+\Lambda(\gamma-3\alpha_i)\partial_iZ\Big)-\frac{1}{3}(W+\Lambda\gamma
Z)^2, \een and the first-order equations read
\ben\label{1st}&&\phi'_i=\pm\frac12\partial_i\Big(W+\Lambda(\alpha_i+\gamma)
Z\Big), \qquad i=1,2,...,N\nonumber\\
&&A'=\mp\frac{1}{3}(W+\Lambda\gamma Z).\een The constraint equations
can be now written as \ben
\partial_i(\partial_iW\partial_iZ)+...+\left[2{\Lambda}(\alpha_i+\gamma)
\partial_i\partial_iZ-\frac{4}{3}(W+{\Lambda}\gamma Z)\right]\partial_iZ=0, \qquad i=1,2,...,N.\een

\section{Gravity Localization}
\label{gravity}
The study of gravity localization on the brane solutions above can
be done by choosing a gauge where the general metric fluctuations
have the form \be ds^2=e^{2 A(y)}(g_{\mu\nu}+\epsilon\;
h_{\mu\nu})dx^\mu dx^\nu-dy^2. \ee Here $g_{\mu\nu}=g_{\mu\nu}(x,y)$
represents the four-dimensional dS, AdS or Minkowski metric, and
$h_{\mu\nu}=h_{\mu\nu}(x,y)$ represents the metric fluctuations, and
$\epsilon$ is a small parameter. Following the
Refs.~\cite{f,kr,c1,c2}, introducing the $z$-coordinate, in order to
turn the metric conformally flat, with $dz=e^{-A(y)}dy$, the metric
fluctuations of the brane solutions, under the choice of transverse
and traceless gauge, leads to the Schroedinger-like equation
\be\label{sch} -\frac{d^2\psi(z)}{dz^2}+U(z)\psi(z)=m^2\psi(z), \ee
with \be\label{Uz} U(z)=\frac94 A^{\prime 2}(z)+\frac32
A^{\prime\prime}(z), \ee for dS, AdS or Minkowski geometry. Note
that this equation can be factorized as \be
\left[-\frac{d}{dz}+\frac34
A^{\prime}(z)\right]\left[\frac{d}{dz}+\frac34
A^{\prime}(z)\right]\psi(z)=m^2\psi(z). \ee This shows that there
are no graviton bound-states with negative mass, and the graviton
zero mode $\psi_{0}(z)=e^{-\frac34 A(z)}$ is the ground-state of the
quantum mechanical problem.

Before going into an explicit example several comments are in order.
Due to difficulty in obtaining $A(z)$ from $A(y)$ in some brane
solutions we have previously considered the calculation of the
graviton spectrum on the brane may require numerical computations.
For example, the study of the metric fluctuations via
Eq.~(\ref{sch}) of the brane solutions defined by the
superpotentials (\ref{wsech}), Eq.~(\ref{wphi4}), and
Eq.~(\ref{wsinh1}),  is only tractable numerically. The case defined
by Eq.~(\ref{ay2}) has been already done in the paper \cite{bg}, for
$a=b=1$. The model defined by the superpotential (\ref{wsinh2})
leads to a Schroedinger-like equation (\ref{sch}) whose potential is
the modified Poschl-Teller type potential, and was studied in
Ref.~\cite{g2,abl2006}. The two-field model given by the
superpotential (\ref{wsinsinh}) is only tractable numerically.
However, for $b=0$ it reduces to the one-field model
$W=3c\sinh(d\chi)$, that for $d=\pm\sqrt{1/3}$ leads to a
volcano-like potential, whose spectrum has been discussed in
Refs.~\cite{f,g1,g2,c1,c2}, and for $d=\pm\sqrt{2/3}$, the spectrum
was investigated in Refs.~\cite{g2,abl2006}. The case $d=0$ gives
the model $W=3a\sin(b\phi)$ whose spectrum is tractable analytically
for $b=\pm\sqrt{1/3}$ and $b=\pm\sqrt{2/3}$ --- see
Ref.~\cite{g1,g2}. The model defined by the superpotential
(\ref{wsincos}) gives $A(y)$ of Eq.~(\ref{ay5a}) and
Eq.~(\ref{ay5b}). The case (\ref{ay5a}), leads to the same situation
of Eq.~(\ref{ay4}) for $d=0$, whereas for the case Eq.~(\ref{ay5b}),
the gravity fluctuations are only tractable numerically.

Finally, we consider explicitly the two-field model for
$\Lambda\neq0$ with the superpotential (\ref{wsin}) that can be
studied analytically. From the $AdS_4$ solution Eq.~(\ref{ay6}) we
have \be A(z)=-\frac12\ln\left(\alpha
a^2\cos^2(\frac{z}{\sqrt{\alpha}})\right). \ee The Schroedinger-like
potential (\ref{Uz}) is now given by \be\label{Uz2}
U(z)=-\frac{9}{4\alpha}+\frac{15}{4\alpha}
\sec^2\left(\frac{z}{\sqrt{\alpha}}\right). \ee This is a
Poschl-Teller potential. The model supports an infinity of bound
states with eigenvalues given by \be\label{spectrum}
m_n^2=\frac{n}{\alpha}\left(n+3\right), \qquad n=1,2,3,... \ee where
$\alpha=(-\Lambda)^{-1}[2/3(b^2+c^2)+1],\, \Lambda<0$. This
potential is the same as the potential found in Karch-Randall
scenario \cite{kr}. The spectrum consists of massive graviton modes
of the gravity fluctuations of a pure $AdS_5$ spacetime. The gravity
localization on the 3-brane is due to a very light mode that appears
as the brane tension becomes sufficiently large
\cite{kr,schwartz,miemiec,bbg,lykken}. Although the brane solution
of the model (\ref{wsin}) has a nonzero tension, e.g.,
$\sigma_{BPS}=|\Delta W|=3a$, for $m=k=0$, such information does not
appear in the potential (\ref{Uz2}). This makes impossible to
control the spectrum (\ref{spectrum}) in order for the lightest
graviton mode responsible for 4d gravity to emerge. If on one hand
it is hard to localize gravity with this brane solution, on the
other hand, as we investigate the RG group flow later, it shows to
be a nice {\it gravity dual} of a weakly coupled field theory on the
$AdS_5$ boundary.

\section{RG flow equations}
\label{RGflow}

According to gauge/gravity duality conjecture such as AdS/CFT
correspondence \cite{malda,gubser,witten} or domain wall/QFT
correspondence \cite{skenderis} there exists the possibility of
considering the warp factor of a spacetime geometry as a scale of
energy of a holographically dual field theory on its boundary. In
this section we are going to consider such conjecture by exploring
the renormalization group flow of the dual field theory. Let us
write the geometry (\ref{metric}) as \ben\label{metric2}
ds^2_5=U^2(y)ds^2_4-dy^2, \een where $U(y)\!=\!e^{A(y)}$. As such,
the warp factor is identified with the renormalization scale $U$ on
the flow equations \cite{s,skenderis,freedman_pilch,dallagata}.

We first consider the case of a single scalar field. We write \ben
\label{rg}
\phi'=\frac{d\phi}{dy}=\frac{dU}{dy}\frac{d\phi}{dU}=A'U\frac{d\phi}{dU}.\een
If the scalar fields on the gravity side is conjectured to be
related to running couplings on the dual field theory side we can
use Eq.~(\ref{rg}) to construct the following beta function
\ben\label{beta} \beta(\phi)\equiv
U\frac{d\phi}{dU}=\frac{\phi'}{A'}=-\frac{3}{2}\frac{W_\phi
+\Lambda(\alpha+\gamma)Z_\phi}{W+\Lambda\gamma Z},\een where we have
used Eqs.~(\ref{A2}) and (\ref{phi2}). Note that the beta function
(\ref{beta}) works for both flat and bent branes supported by a
single scalar field. At critical points $\phi=\phi^*$ (or
$\phi=\phi_{vac}$ for supersymmetric vacua) the beta function
vanishes. Thus, expanding the beta function $\beta(\phi)$ around the
critical point we find
\ben\label{beta_p}\beta(\phi)=\beta(\phi^*)+\beta'(\phi^*)(\phi-\phi^*)+
..., \een where $\beta(\phi^*)=0$, and $\beta'(\phi^*)$ can be
expressed in terms of $W$ and $Z$ as
\ben\label{beta_p2}\beta'(\phi^*)=-\frac32\left[\frac{W_{\phi\phi}
+\Lambda(\alpha+\gamma)Z_{\phi\phi}}{W+\Lambda\gamma
Z}-\frac{\big(W_\phi+\Lambda(\alpha+\gamma)Z_\phi\big)\big(W_\phi+\Lambda\gamma
Z_\phi\big)}{(W+\Lambda\gamma Z)^2}\right]_{\,\phi=\phi^*}. \een
Combining the equations (\ref{beta}) and (\ref{beta_p}) and
integrating out both sides one can find the following running
coupling equation
\ben\label{coup}\phi=\phi^*+c\,U^{\,\beta'(\phi^*)}, \een where $c$
is a constant. For $\beta'(\phi^*)<0$ and energy scale $U\to\infty$
we have that $\phi=\phi^*$ is an ultraviolet (UV) stable {\it fixed
point}, whereas for $\beta'(\phi^*)>0$ and energy scale $U\to0$ we
have that $\phi=\phi^*$ is an infrared (IR) stable {\it fixed
point}. An $AdS_5$ vacuum solution $U=e^{ky}$ gives a ( weak )
strong running coupling $\phi$ as $y\to\infty$, i.e., $U\to\infty$,
for ( $\beta'(\phi^*)<0$ ) $\beta'(\phi^*)>0$.

Let us investigate our brane solutions whose kink profile can be
identified with running coupling of the dual field theory. For the
$Z_2$-symmetric branes $U(y\to+\infty)=U(y\to-\infty)$ such that it
is enough to focus just on one slice of the 5d spacetime, say,
$U=e^{A(y)},\,(y>0)$ at the vacuum ($y\to\infty$). It is
interesting, from the AdS/CFT correspondence, those solutions which
are asymptotically $AdS_5$ with $U(y\to\infty)=\infty$ (UV stable
fixed point).

The $\lambda\phi^4$ example with $\Lambda=0$ in (\ref{wphi4}) gives
us $\beta'(\phi^*)=9b^2/2$ which means there exists an IR stable
fixed point on the dual field. This result signalizes gravity
localization on the brane \cite{kallosh_linde,cvetic2000}.

On the other hand, the bent brane example with $\Lambda<0$ in
(\ref{wsinh1}) gives us $\beta'(\phi^*)=-3b^2/2$ that implies the
existence of an UV stable fixed point on the dual field.  Thus, this
field theory is a weakly coupled theory at high energy, although the
coupling never diverges because the kink smoothly connects two
different vacua  with the same scale $U(y=\pm\infty)=\infty$. As a
comparison, recall for the dilaton domain wall
\cite{skenderis,freedman_pilch}, $U(y=-\infty)=0$ and
$U(y=\infty)=\infty$. In our bent brane solution we find that at
$y=0$, the smallest distance in the bulk at one side of the brane,
the energy scale becomes $U=1$. This signalizes that a non-confining
phase in the infrared regime may appear. There is a ``natural'' IR
cut-off in this space. Note that asymptotically, i.e.,
$y\to\pm\infty$, $A(y)\simeq \pm y/R$, one has $AdS_5$ slices.  Now
changing the coordinates of the metric (\ref{metric2}) as
$U=e^{A(y)}=r/R$, one finds
\ben\label{metric3}ds_5^2=\frac{r^2}{R^2}\,ds_4^2-\frac{R^2}{r^2}\,dr^2.
\een These $AdS_5$ slices are connected by the $AdS_4$ brane at
$y=0$. Of course, the range of the coordinate $r$ for a slice, say,
$y\geq0$, is restricted to $R\leq r<\infty$, such that $r=R$ is an
infrared cut-off of this $AdS_5$ slice. Thus the position of the
$AdS_4$ brane at $y=0$ (or equivalently at $r=R$) is a natural
infrared cut-off. In  recent developments \cite{braga,herzog}, in
which one considers the introduction of IR cut-off to obtain a
deconfining phase transition, one extends an $AdS_5$ metric like the
metric (\ref{metric3}) to an AdS-Schwarzschild metric in
ten-dimensions, whose deconfining temperature is given in terms of a
relation between the horizon radius and the infrared cut-off.

Note also that the brane in the case $\Lambda>0$, because of its
singular behavior, does not present a well defined beta function
--- same happens to the model (\ref{wsinh2}). As we have earlier
discussed, the negative cosmological constant also resolves the
singularity of the brane at infrared regime --- see
Fig.~\ref{fig5c}.

Another interesting example is the one given in (\ref{wsech}), whose
kink profile connects vacua at infinity. It produces
$\beta'(\phi^*)=0$, which means from (\ref{coup}) that $\phi=\phi_*$
is fixed everywhere, that is, we have on the boundary, a  dual
conformal field theory.

The extension to a theory with multi-running couplings $\phi^i$ is
straightforward. The equations (\ref{beta}) and (\ref{beta_p}) can
be now combined in the form
\ben\label{beta_p2multi}\beta^i(\phi)\equiv
U\frac{d\phi^i}{dU}=\frac{\partial\beta^i(\phi^*)}{\partial\phi^j}(\phi^j-{\phi^j}^*)+...\een
where $\beta^i(\phi^*)$ is Eq.~(\ref{beta_p2}) for multi-running
couplings and ${\partial\beta^i(\phi^*)}/{\partial\phi^j}$ the
corresponding derivatives. Now we are ready to discuss kink profiles
of branes found in the two-field models that we have considered
earlier.

The model $W=3a\sin(b\phi)\cos(b\chi)$ with $\Lambda=0$ -- see
Eq.~(\ref{wsincos}) -- has the following derivative of the beta
functions
\ben\label{beta_p2phi_chi}
\frac{\partial\beta^\phi(\phi^*)}{\partial\phi}=\frac{3}{2}b^2,\,\,\,
\frac{\partial\beta^\chi(\phi^*)}{\partial\chi}=\frac{3}{2}b^2,\,\,\,\,
\frac{\partial\beta^\phi(\phi^*)}{\partial\chi}=\frac{\partial\beta^\chi(\phi^*)}{\partial\phi}=
-\frac{3}{2}b^2\tan^2{\left(\frac{k\pi}{2}\right)}
\een
This was done for $C=1$. In the solution for $C=0$, the second
derivative adds to zero. Since $k$ are integer numbers, the second
derivative above is finite only if $k$ is $even$. This is the case
of the $asymmetric$ supersymmetric $\mathbb{M}_5-AdS_5$ brane
discussed earlier. The running couplings equations are
\ben\phi^i={\phi^i}^*+c^iU^{\frac{3}{2}b^2},\qquad
\phi^i=(\phi,\chi).
\een
In this model $U(y\to\infty)\to0$ such that
we have an IR stable fixed point $\phi^i={\phi^i}^*$. Note also that
on the Minkowski side of the brane $U(y=-\infty)=const<\infty$ --
see Fig.~\ref{fig8}--, thus the running couplings have a {\it
`natural' UV cut-off}. The running couplings $(\phi,\chi)$ vary
their strengths in the same way.

Let us now finish this section by discussing the beta function of
the ``bent'' brane model $W=3a\sin(b\phi+c\chi)$ with $\Lambda\neq0$
and $Z=W$ -- see Eq.~(\ref{wsin}). For regular kink profile the
cosmological constant $\Lambda<0$ and supersymmetric vacua satisfy
the relation $1+\Lambda\alpha>0,\, \alpha>0$. The beta functions in
this case obey
\ben\label{beta_p3phi_chi}
\frac{\partial\beta^\phi(\phi^*)}{\partial\phi}=\frac{3}{2}b^2(1+\Lambda\alpha),
\,\,\,
\frac{\partial\beta^\chi(\phi^*)}{\partial\chi}=\frac{3}{2}c^2(1+\Lambda\alpha).
\een
and
\be\label{beta_p3phi_chi1}
\frac{\partial\beta^\phi(\phi^*)}{\partial\chi}=\frac{\partial\beta^\chi(\phi^*)}{\partial\phi}=
\frac{3}{2}bc(1+\Lambda\alpha),
\ee
The equations (\ref{beta_p2multi}) becomes
\ben\label{beta_p4phi}U\frac{d\phi}{dU}=
\frac{\partial\beta^\phi(\phi^*)}{\partial\phi}(\phi-{\phi}^*)+
\frac{\partial\beta^\phi(\phi^*)}{\partial\chi}(\chi-{\chi}^*)+...\\
\label{beta_p4chi}
U\frac{d\chi}{dU}=\frac{\partial\beta^\chi(\phi^*)}{\partial\phi}(\phi-{\phi}^*)
+\frac{\partial\beta^\chi(\phi^*)}{\partial\chi}(\chi-{\chi}^*)+...
\een
Substituting the explicit beta functions (\ref{beta_p3phi_chi}) and (\ref{beta_p3phi_chi1})
into Eqs.~(\ref{beta_p4phi}) and (\ref{beta_p4chi}), summing up one
another and integrating out the resulting equation one can find that
the running couplings $\phi^i=(\phi,\chi)$ vary according to the
formula \ben\phi^i={\phi^i}^*+\frac{c^i}{U},\een where we have used
the fact that $b^2+c^2=-2/3(1+\Lambda\alpha)$. Again, we have here a
dual field theory exhibiting an weakly coupled regime at high
energy. The running couplings $\phi, \chi$, are fixed, i.e.
$\phi^i={\phi^i}^*$, as $U=e^{A}\to\infty$ in the slice $y>0$ of the
$AdS_4$ brane that we have found earlier.

Let us now return to the discussion about gravity localization for
this brane solution. Our former calculation shows that there exist
only {\it massive gravity} on the spectrum. The gravity localization
is favored as long as a very light graviton mode emerges, such that
gravity is locally localized \cite{kr,schwartz,miemiec,bbg,lykken},
although asymptotically the warp factor diverges. This agrees with
the well-known fact that there is no normalizable graviton zero mode
as the warp factor diverges
\cite{rs2,c1,c2,kallosh_linde,cvetic2000,kr,bbg,lykken}.

\section{Ending comments}
\label{conclu}

In this work we have shown how to write a first-order  formalism to
braneworld scenarios which include the possibilities of the brane to
have $AdS$, Minkowski, and $dS$ geometry. The crucial ingredient was
the introduction of two new functions, $W=W(\phi)$ and $Z=Z(\phi)$
from which we could express both $A$ and $\phi$ in terms of
first-order differential equations, for the potential engendering
very specific form. The importance of the procedure is related not
only to the improvement of the process of finding explicit solution,
but also to the opening of another route, in which we can very fast
and directly write the warp factor once $W(\phi)$ and $Z(\phi)$ are
given. As we have shown, the present investigations seem to open
several distinct possibilities of study.

The issue concerning the gauge/gravity duality is interesting, and
the first-order formalism here developed for $AdS$, Minkowski, or $dS$
geometry can easily be used to find the renormalization
group flow of the holographically dual
field theory. By considering some models given in terms of specific
superpotentials, we have shown that there are interesting ``bent''
brane solutions with negative cosmological constant ($AdS_4$ branes)
that can play the role of nice gravity duals. The RG flow shows that
UV stable fixed point of a dual field theory can be treated
perturbativelly at high energy, and just like QCD it may develop
`asymptotic freedom'. Such ``bent'' branes may give a dual
gravitational description of RG flows in supersymmetric field
theories living in the curved spacetime of the brane world-volume
\cite{dallagata}. Furthermore, as we have shown in an explicit
example, $AdS_4$ branes seem to exhibit improved infrared behavior.
Other examples include {\it asymmetric branes} that are
asymptotically $\mathbb{M}_5-AdS_5$ spaces. We found that in these
theories there exists a `natural' UV cut-off on the running coupling
in the Minkowski ($\mathbb{M}_5$) side. All the brane solutions we
found here are connecting two different vacua. A natural
continuation is to look for solution connecting critical points
other than vacua, such as local maxima, saddle points, and so on
\cite{freedman_pilch} to look for other gravity duals.

Evidently, in addition to the subject of gravity localization on
thick branes, the interest on the subject broadens with the
application of the method to investigate new gravity duals and
renormalization group flow of the holographically dual field theory,
since now we can easily find the flow equations for brane models
with arbitrary cosmological constant and bulk scalar fields that can
be identified with running couplings in the dual field theory.

\acknowledgments

The authors would like to thank V.I. Afonso and R. Menezes for discussions,
and CAPES, CNPq, and PRONEX/CNPq/FAPESQ for partial support.



\begin{thebibliography}{99}

\bb{rs2}L. Randall and R. Sundrum, Phys. Rev. Lett. {\bf83,} 4690
(1999); [arXiv:hep-th/9906064].

\bb{gw}W.D. Goldberger and M.B. Wise, Phys. Rev. Lett. {\bf83,} 4922
(1999); [arXiv:hep-ph/9907447].

\bb{s}K. Skenderis and P.K. Townsend, Phys. Lett. B {\bf468,} 46
(1999); [arXiv:hep-th/9909070].

\bb{f}O. DeWolfe, D.Z. Freedman, S.S. Gubser, and A. Karch, Phys.
Rev. D {\bf62,} 046008 (2000); [arXiv:hep-th/9909134].

\bb{g1}M. Gremm, Phys. Lett. B {\bf478,} 434 (2000);
[arXiv:hep-th/9912060].

\bb{g2}M. Gremm, Phys. Rev. D {\bf62,} 044017 (2000);
[arXiv:hep-th/0002040].

\bb{c1}C. Cs\'aki, J. Erlich, T.J. Hollowood, and Y. Shriman, Nucl.
Phys. B {\bf581,} 309 (2000); [arXiv:hep-th/0001033].

\bb{c2}C. Cs\'aki, J. Erlich, C. Grogean, and T.J. Hollowood, Nucl.
Phys. B {\bf584,} 359 (2000); [arXiv:hep-th/0004133].

\bb{klp}N. Kaloper, Phys. Rev. D {\bf60}, 123506 (1999);
[arXiv:hep-th/9905210].

\bb{kr} A. Karch, L. Randall, JHEP {\bf0105},  008 (2001);
[arXiv:hep-th/0011156].

\bb{bfg}D. Bazeia, C. Furtado, and A.R. Gomes, JCAP {\bf0402,} 002 (2004).

\bb{fn}D.Z. Freedman, C. N\'u\~nez, M. Schnabl, and K. Skenderis,
Phys. Rev. D {\bf69,} 104027 (2004); [arXiv:hep-th/0312055].

\bb{celi}A. Celi, A. Ceresole, G. Dall'Agata, A. Van Proeyen, M.
Zagermann, Phys. Rev. D {\bf71,} 045009 (2005);
[arXiv:hep-th/0410126].

\bb{bcy}F.A. Brito, M. Cvetic, S.-C. Yoon, Phys. Rev. D64 (2001)
064021; [arXiv:hep-ph/0105010].

\bb{cv_lamb} M. Cvetic, N.D. Lambert, Phys. Lett. B {\bf540}, 301
(2002); [arXiv:hep-th/0205247].

\bb{bbn}D. Bazeia, F.A. Brito, J.R. Nascimento, Phys. Rev. D {\bf68}
085007 (2003); [arXiv:hep-th/0306284].

\bb{bglm}D. Bazeia, C.B. Gomes, L. Losano, and R. Menezes, Phys.
Lett. B {\bf633,} 415 (2006); [arXiv:astro-ph/0512197].

\bb{abl2006}V.I. Afonso, D. Bazeia and L. Losano, Phys. Lett. B 
{\bf634}, 526 (2006); [arXiv:hep-th/0601069].

\bb{st}K. Skenderis and P.K. Townsend, Phys. Rev. Lett. {\bf96,}
191301 (2006); [arXiv:hep-th/0602260].

\bb{st2}K. Skenderis and P.K. Townsend, {\it Hamilton-Jacobi for
domain walls and cosmologies}; [arXiv:hep-th/0609056].

\bb{skenderis}H.J. Boonstra, K. Skenderis, P.K. Townsend, JHEP {\bf9901,}
003 (1999); [arXiv:hep-th/9807137].

\bibitem{freedman_pilch} D.Z. Freedman, S.S. Gubser, K. Pilch,
and N.P. Warner, Adv. Theor. Math. Phys. {\bf3}, 363 (1999);
[arXiv:hep-th/9904017].

\bb{dallagata}G. L. Cardoso, G. Dall'Agata, D. Lust, JHEP {\bf0203,}
 044 (2002); [arXiv:hep-th/0201270].

\bb{V}I. Cho and A. Vilenkin, Phys. Rev. D {\bf59,} 021701(R)
(1999); [arXiv:hep-th/9808090]; Phys. Rev. D {\bf59,} 063510 (1999);
[arXiv:gr-qc/9810049].

\bb{p}D. Bazeia, Phys. Rev. D {\bf60,} 067705 (1999);
[arXiv:hep-th/9905184].

\bb{wang} A. Wang, Phys. Rev. D {\bf66},  024024 (2002);
[arXiv:hep-th/0201051].

\bb{melfo1}A. Melfo, N. Pantoja, A. Skirzewski, Phys. Rev. D {\bf67},
105003 (2003); [arXiv:gr-qc/0211081].

\bb{melfo2}O. Castillo-Felisola, A. Melfo, N. Pantoja, A. Ramirez,
Phys. Rev. D {\bf70}, 104029 (2004); [arXiv:hep-th/0404083].

\bb{ts}K. Takahashi and T. Shiromizu, Phys. Rev. D {\bf70,} 103507
(2004); [hep-th/0408043].

\bb{pad}A. Padilla, Class. Quant. Grav. {\bf22,} 681 (2005);
[arXiv:hep-th/0406157].

\bb{gort}R. Guerrero, R. Omar Rodriguez, and R. Torrealba, Phys.
Rev. D {\bf72,} 124012 (2005); [arXiv:hep-th/0510023].

\bb{cvetic92}M. Cvetic, S. Griffies, S.-J. Rey, Nucl. Phys. {\bf
B381}, 301 (1992); [arXiv:hep-th/9201007].

\bb{cvetic93}M. Cvetic, S. Griffies, H. H. Soleng, Phys. Rev.
D {\bf48}, 2613 (1993);[arXiv:gr-qc/9306005].

\bb{cvetic97} M. Cvetic, H. H. Soleng, Phys. Rept. {\bf282}, 159
(1997); [arXiv:hep-th/9604090].

\bb{bsr}D. Bazeia, M.J. dos Santos, R.F. Ribeiro, Phys. Lett.
A {\bf208}, 84 (1995); [arXiv:hep-th/0311265].

\bb{bbet}J.D. Edelstein, M.L. Trobo, F.A. Brito, and D. Bazeia,
Phys. Rev. D {\bf57}, 7561 (1998); [arXiv:hep-th/9707016].

\bb{bg}D. Bazeia, A.R. Gomes, JHEP {\bf0405,} 012 (2004);
[arXiv:hep-th/0403141].

\bb{schwartz}M.D. Schwartz, Phys. Lett. B {\bf502}, 223 (2001);
[arXiv:hep-th/0011177].

\bb{miemiec}A. Miemiec, Fortsch. Phys. {\bf49}, 747 (2001);
[arXiv:hep-th/0011160].

\bb{bbg}D. Bazeia, F.A. Brito, A.R. Gomes, JHEP {\bf0411}, 070
(2004); [arXiv:hep-th/0411088].

\bb{lykken}R. Bao, M. Carena, J. Lykken, M. Park, and J. Santiago,
Phys. Rev. D {\bf73}, 064026 (2006); [arXiv:hep-th/0511266].

\bb{malda}J.M. Maldacena, Adv. Theor. Math. Phys. {\bf2}, 231
(1998); [arXiv:hep-th/9711200].

\bb{gubser}S.S. Gubser, I.R. Klebanov, A.M. Polyakov, Phys. Lett.
B {\bf428}, 105 (1998); [arXiv:hep-th/9802109].

\bb{witten}E. Witten, Adv. Theor. Math. Phys. {\bf2}, 253 (1998);
[arXiv:hep-th/9802150].

\bb{kallosh_linde}R. Kallosh, A. Linde, JHEP {\bf0002}, 005 (2000);
[arXiv:hep-th/0001071].

\bb{cvetic2000} M. Cvetic, Int. J. Mod. Phys. A {\bf16}, 891 (2001);
[arXiv:hep-th/0012105].

\bb{braga}H. Boschi-Filho, N.R.F. Braga, C.N. Ferreira, ``Heavy
quark potential at finite temperature from gauge/string duality'';
[arXiv:hep-th/0607038].

\bb{herzog}  C.P. Herzog, ``A holographic prediction of the
deconfinement temperature''; [arXiv:hep-th/0608151].

\end{thebibliography}
\end{document}